\documentclass[final,authoryear,3p,times,twocolumn]{elsarticle}

\usepackage{graphicx}
\usepackage{caption}
\usepackage{subcaption}
\usepackage{todonotes}
\usepackage{placeins} % provides \FloatBarrier

\usepackage{float}
\usepackage{fixltx2e}
\usepackage{amsmath}

\usepackage{amssymb}

\usepackage[breaklinks=true,pdftex]{hyperref} %% to avoid \citeads line fills

\graphicspath{{images/}}

\def\arcmin{\hbox{$^\prime$}}

\newcommand{\degree}{\ensuremath{^\circ}\,}
\newcommand{\degreenospace}{\ensuremath{^\circ}}

\journal{Astronomy and Computing}

\begin{document} 

\begin{frontmatter}

   \title{Destriping Cosmic Microwave Background Polarimeter data}

\author[ucsb]{A. Zonca\corref{cor1}}
\ead{zonca@deepspace.ucsb.edu}

\author[ucsb]{B. Williams}
\ead{brianw@deepspace.ucsb.edu}

\author[ucsb]{P. Meinhold}
\ead{peterm@deepspace.ucsb.edu}

\author[ucsb]{P. Lubin}
\ead{lubin@deepspace.ucsb.edu}

\cortext[cor1]{Corresponding author}

\address[ucsb]{Department of Physics, University of California, Santa Barbara\\
              Santa Barbara, CA 93106 \\
}

\date{\today}
 
\begin{abstract}
Destriping is a well-established technique for removing low-frequency correlated noise from Cosmic
Microwave Background (CMB) survey data.  In this paper we present a destriping algorithm tailored to
data from a polarimeter, i.e.\ an instrument where each channel independently measures the
polarization of the input signal.  

We also describe a fully parallel implementation in Python
released as Free Software and analyze its results and performance on simulated datasets,
both the design case of signal and correlated noise, 
and with additional systematic effects.

Finally we apply the algorithm to 30 days of 37.5 GHz polarized microwave data gathered from the
B-Machine experiment, developed at UCSB\@. The B-Machine data and destriped maps are made publicly
available.  

The purpose is the development of a scalable software tool to be applied to the upcoming
12 months of temperature and polarization data from LATTE (Low frequency All sky TemperaTure
Experiment) at 8 GHz and to even larger datasets.
\end{abstract}


\end{frontmatter}
             
\section{Introduction}

The destriping technique \citep{kurki09, keihanen09, ashdown07, delab98, maino02, natoli01,
    revenu00,
poutanen06} is widely used in the CMB field for removing correlated low-frequency noise, typically
due to thermal drifts, amplifiers gain fluctuations and changes in atmospheric emission.  To be
effective, destriping requires that the data are acquired with a scanning strategy \citep{dupac05}
which assures frequent crossing points, i.e.\ looking at the same sky location at different times.

Correlated noise at low frequency is often modeled as $1/f$ noise:

\begin{equation} 
    P(f) \propto \sigma^2 \left[ 1 + \left( \dfrac{f_k}{f} \right)^\alpha \right], \label{eq:1fnoise}
\end{equation} where: 

\begin{itemize} 
    \item $f$ is the frequency, 
    \item the knee frequency $f_k$ denotes the frequency where white and correlated noise
            contribute equally to the total noise 
    \item the power spectrum below the knee frequency
        is increasing as a power law with exponent $\alpha$. 
\end{itemize}

For a typical survey experiment, making multiple passes over the same region of sky, 
destriping can efficiently remove 1/f noise with a knee frequency lower than the scan rate 
by removing a sequence of offsets (called baselines) of a predetermined length, typically of order the scan length (or spin period). 
The baselines are estimated iteratively by binning the data on a sky map and minimizing 
the scatter among different measurements of each pixel.

The destriping algorithm allows the removal of
the correlated noise with minimal impact on the signal \citep{efstathiou05, maino99}. Destriping can
be thought of as a high-pass filter that only affects the noise in the timestream.

This paper first introduces (Sec.~\ref{sec:polarimeter}) the B-Machine experiment \citep{williams10}
to highlight the features of a CMB polarimeter and the typical data processing required to convert
raw data to calibrated timelines ready for map-making and noise characterization.
Sec.~\ref{sec:algorithm} describes in detail the destriping technique, a Maximum Likelihood
algorithm for removing low frequency correlated noise from the data. We describe in detail the
implementation specifics in Sec.~\ref{sec:implementation} for application of destriping to a polarimeter and present results of the
destriping on simulated data in Sec.~\ref{sec:simulations}.  Sec.~\ref{sec:bmachine} is dedicated to the application of destriping
to B-Machine data, analyzing the impact of destriping in map, frequency and angular power spectrum
domain.

The B-Machine experiment itself and the data analysis pipeline presented in this work are focused 
on preparing the upcoming LATTE (Low frequency All sky TemperaTure
Experiment) experiment.

LATTE is a ground based telescope designed to survey the
diffuse microwave foregrounds, primarily from the Milky
Way.  Its main purpose is to help fill the gap in the
data that exists between the 408 MHz Haslam \citep{haslam82} survey and the
lowest WMAP band at 23 GHz \citep{gold11}.  The data from LATTE
will be useful for the study of the CMB by better
characterizing foreground emission so that they can be
more effectively removed from WMAP and Planck \citep{planckcompsep13} data.  But it will
also add to our understanding of the Inter Stellar Medium (ISM) \citep{spitzer08},
in particular spinning dust emission \citep{planckdust11} and Galactic haze \citep{dobler08,planckhaze13}.

LATTE will have a single detector with a 30\% bandwidth
centered at 8 GHz (future plans include bands in the 3 to 15 GHz
range) and an angular resolution of about 2\degreenospace, 
it will be based on cryogenically cooled radiometers to 
measure relative sky temperature and polarization
with an expected sensitivity as low as 340 $\mathrm{\mu K \sqrt{s} }$.
For comparison the C-Bass \citep{king10} 5 GHz radiometer, an experiment with similar scientific goals,
has a bandwidth of 1.4 GHz and a sensitivity of about 580 $\mathrm{\mu K \sqrt{s} }$.

\section{Polarimeters data}
\label{sec:polarimeter}

B-Machine is a prototype ground-based CMB polarimeter operated at White Mountain Research Station in
California during 5 months in 2008, see Table~\ref{tab:bmachineperf} for a summary of the experiment
features, and the instrument paper \citep{williams10} for a detailed description.

\begin{table}[htb]
      \caption[]{Summary of B-Machine.}
      \label{tab:bmachineperf}
     $$
         \begin{array}{p{0.4\linewidth}l}
            \hline          
            \noalign{\smallskip}
            Frequency                         & 37.5\,\mathrm{GHz}         \\
            Beam full width half max          & 22.2\,\mathrm{arcmin}           \\
            Number of horns                   & 4 \\
            Scanning strategy                 & 45^\circ\,\mathrm{elevation}, 70\mathrm{s}\, \mathrm{revolution}\\
            Sensitivity ($I$)      & 16, 16.4, 23.1, 5.3\ \mathrm{mK} / \sqrt{\mathrm{Hz}} \\
            Sensitivity ($Q$, $U$)      & 7.6, 6.7, 10.5, 2.8\ \mathrm{mK} / \sqrt{\mathrm{Hz}} \\
            $1/f$ noise knee frequency in $Q$ and $U$     & 5.0\,\mathrm{mHz} \\
            Amount of valid data               & 314,\ 329,\ 30,\ 293\ \mathrm{hours} \\  
            Altitude      & 3800\,\mathrm{m} \\
            Atmosphere temperature     & 10\,\mathrm{K} \\
            \noalign{\smallskip}
            \hline
         \end{array}
     $$ 
   \end{table}
  
The key enabling technology of the B-Machine polarimeter is a reflection half-wave plate rotating at
30 Hz, located between the primary and the secondary mirrors of our 2~m Gregorian telescope.  The wave
plate continuously rotates the polarization of the incoming waves.  Viewing the sky through this
wave plate, the radiometer is sensitive to a linear polarization state which rotates 120 times per
second, and this measurement can be demodulated to completely solve for the linear polarization
state of the incoming radiation.

In detail, for each disk rotation, each polarimeter outputs a measurement for  $q$ and one for $u$,
i.e.\ the $Q$ and $U$ Stokes polarization states in the reference frame of the telescope. These timestreams need then to be rotated to the Equatorial or Galactic $Q$ and $U$ using the known pointing and orientation information.

Demodulation has also the effect of highly suppressing the gain fluctuations of the receiver,
effectively being equivalent to chopping 4 times per revolution, i.e. 120 Hz, with a reference
target.  
Although B-Machine detectors were designed as pure polarimeters, it is possible also to measure the intensity $I$ of the incoming signal by integrating the signal across the disc rotation.
However, without modulating and then demodulating $I$ is strongly affected by correlated noise and
difficult to use for scientific purposes. For the LATTE experiment we are including a high-frequency
switch that allows switching between the sky and a reference target to provide a performance 
comparable to the polarized channels.

A complete absolute calibration with a reference target was performed only twice, 
at the beginning and at the end of the campaign.
Data were calibrated to physical units with a daily relative calibration.

Data were sampled at 33.4 Hz, but, due to bad
samples or operational issues in the instrument, they show several gaps. There are two primary
reasons we decided to remove flagged samples and to fill the gaps with white noise: 

\begin{itemize}
\item Noise characterization requires computing Fast Fourier Transforms (FFTs), which generally
    require constant sampling frequency.  Flagged samples need to be replaced by white noise.  
\item The destriper, for ease of implementation, does not use timing information but just a fixed
        number of samples per offset, i.e. 2000, therefore it assumes a constant sampling rate.
\end{itemize} 

Our white-noise filling algorithm just replaced all flagged and missing samples with
gaussian white noise with the same mean and standard deviation of the good samples for each day of
data.

Pointing reconstruction and calibration outputs consist of about 3 GB of data for each channel,
containing: 

\begin{itemize} 
    \item timing information 
    \item pointing angles 
    \item bad data flags
    \item $I$, $q$ and $u$ measurements calibrated to physical units [K].  
\end{itemize}

\section{Algorithm}
\label{sec:algorithm}

B-Machine temperature and polarization outputs are fully decoupled, because the polarimeter measures
$q$ and $u$ directly, and binning over a full revolution to measure temperature effectively removes
any polarized emission and gives a pure $I$ component.  Map-making can be therefore performed
independently on the temperature and polarization outputs; temperature-only destriping and $IQU$
destriping of polarization-sensitive detectors are already  well described in the literature
\citep{kurki09, ashdown07, delab98, maino02, revenu00, poutanen06}, in the
following section we will focus on the specifics of destriping for polarimeters, i.e.\ $QU$ only
destriping.

Destriping models the output of each of the two channels of a polarimeter as: 

\begin{equation} 
    y = (R_Q P m_Q + R_U P m_U) + n, 
\label{eq:destmodel} 
\end{equation} 

where: 
\begin{itemize} 
    \item $y$ is
        the data timeline, an array of length $n_t$, already calibrated in physical units [K] 
    \item $m_Q$ , $m_U$ are the $Q$ and $U$ component maps pixelized using the
            HEALPix\footnote{HEALPix website:\url{http://healpix.jpl.nasa.gov}} scheme, each of
            length $n_{pix}$ 
    \item $P$ is the pointing matrix, a $[n_t, n_{pix}]$ sparse matrix with
                one element equal to 1 for each row corresponding to the pixel pointed by the
                detector in that sample. Therefore $Pm$ is the scanning of the map to a timeline,
                given the pointing embodied in $P$. The pointing matrix is the same for both
                channels of each horn, 
    \item $n$ is the white plus correlated noise timeline.
\end{itemize} 

For polarization sensitive detectors, it is also necessary to rotate the
polarized $QU$ components in the detector reference frame.  This is accomplished by the
diagonal $[n_t, n_t]$ matrices $R_Q$, $R_U$ where each element takes into account the
current orientation of the detectors.  Given the measured parallactic angle\footnote{The parallactic angle $\psi$ is the angle between the direction of
polarization sensitivity of the detector and the meridian passing by the current
pointing direction}
$\psi$, $R_Q$ and $R_U$ definition depends whether the channel is $q$ or $u$
polarization: 

\begin{align} 
    q\,\mathrm{channel}:  R_Q[t,t] &= \cos(2 \psi[t]) \\ 
                          R_U[t,t] &= \sin(2 \psi[t]) \\ 
    u\,\mathrm{channel}:  
                          R_Q[t,t] &= -\sin(2 \psi[t])\\ 
                          R_U[t,t] &= \cos(2 \psi[t]), 
\end{align}
   
where $t$ is the sample index or sample time.  Of course in the implementation those diagonal
matrices can just be stored as an array of length $n_t$ and multiplied element-wise to the $Pm$
timeline.

The noise timeline can be separated in two components: 

\begin{equation} 
    n = Fa + w, 
    \label{eq:noise}
\end{equation} 

a white noise component $w$ and a correlated component modeled by the baselines
array $a$ of length $n_b$ multiplied by the $[n_t, n_b]$ sparse matrix $F$. Each row of $F$ has only
one nonzero element, equal to 1, at the column corresponding to the baseline which includes the
current sample. Therefore the matrix repeats each of the elements in $a$ by the baseline lengths and
outputs an array of length $n_t$.

Substituting Eq.~\ref{eq:noise} in Eq.~\ref{eq:destmodel} the complete model is: 

\begin{equation} 
    y = (R_Q P m_Q + R_U P m_U) + Fa + w.  
\end{equation}

First we can compute the hitmap, the map which contains the number of hits for each pixel, as:
\begin{equation} h = P^T P. \end{equation}

Using Maximum Likelihood analysis \citep{kurki09}, we can compute the map $m$, given the baselines
$a$: 

\begin{align} 
    m_Q &= h^{-1} P^T \left[R_{Qq} (y_q - Fa_q) + R_{Qu} (y_u - Fa_u)\right] \label{eq:mapQ} \\ 
    m_U &= h^{-1} P^T \left[R_{Uq} (y_q - Fa_q) + R_{Uu} (y_u - Fa_u)\right].\label{eq:mapU} 
\end{align} 

In this equation, first we remove the $1/f$ noise from
$y_q$ and $y_u$ by subtracting the given baselines, then the signal is projected by the $R$ matrices
to the sky reference frame.  Noise weighting is not necessary because we are building each horn map
independently, and we assume noise stationarity.  Then the $P_T$ matrix sums all the observations
into the two maps, typically called sum maps, which are then divided by the hitmap $h$ to average
the observations.

In $QU$ polarimeter map-making, each pixel is by design measured simultaneously by the $q$ and $u$
channels, therefore the weighting process is simpler than in the polarization-sensitive radiometer
$IQU$ map-making case where instead it is necessary to invert a $[3, 3]$ matrix.

Both Eq.~\ref{eq:mapQ} and Eq.~\ref{eq:mapU} can be rewritten by merging the binning operation in a
single matrix $B$: 

\begin{equation} m = B ( y - Fa ) \end{equation} 

where: 
\begin{align*} 
                B &= h^{-1} \tilde{P}^TR\\ 
    \tilde{P} &=  \begin{bmatrix}P & \\ 
                  & P\end{bmatrix}\\ 
                R &= \begin{bmatrix}R_{Qq} & R_{Uq}\\ 
           R_{Qu} & R_{Uu}\end{bmatrix} 
\end{align*}

Replacing Eq.~\ref{eq:mapQ} and Eq.~\ref{eq:mapU} in the Maximum Likelihood equation and solving
instead for the baselines, we get the destriping equation: 

\begin{equation} 
    \left[ \tilde{F}^T
    \left( I - R\tilde{P}B \right) \tilde{F} \right] a = \tilde{F}^T \left( I -
    R\tilde{P}B \right) y 
\end{equation} 
    
where: 
\begin{align*} 
    \tilde{F} &= \begin{bmatrix}F & \\ 
                  & F\end{bmatrix}\\
    y             &= \begin{bmatrix}y_q \\ 
    y_u\end{bmatrix}\\ 
    a &= \begin{bmatrix}a_q \\
    a_u\end{bmatrix}.  
\end{align*} 

The matrix $R\tilde{P}B$ rescans
the binned map created by $B$ to a timeline, therefore the matrix $I-R\tilde{P}B$ removes the
        estimated signal from the timeline, and then the $F^T$ matrix bins it for each baseline
        period.  The right hand side of the destriping equation performs signal removal on the input
        data $y$, and the residual is summed over for each baseline period; the final result is an
        array of length $n_b$ that contains the residual noise averaged by baseline.
        The left hand side performs exactly the same operation but on the timeline $\tilde{F}a$,
        i.e.\ the timeline composed just of the baselines, which are the unknowns in this equation.

The purpose of a destriper is to find the baseline array that produces the same residual noise of
the data over each baseline period.  The left-side matrix of the destriping equation is a dense
matrix of size $[n_b, n_b]$, and even for modest datasets it is computationally too heavy to be
inverted, and typically it is never explicitely computed, either; iterative solvers like Conjugate
Gradient (CG) \citep{shewchuk94} or Generalized Minimal RESidual method (GMRES) \citep{saad86gmres} allow  to estimate the baselines with
the required accuracy by just applying algorithmically the steps needed to apply the left-hand
operator, without involving any large matrix operation.

\section{Implementation}
\label{sec:implementation}

B-Machine itself is a prototype experiment, built to test the design of the polarization rotator and
to explore the challenges of a ground based polarimeter.  A similar approach has been taken for the
software design, the purpose is to implement a data analysis pipeline that works for the current
dataset but is ready to be extended for future experiments, where the amount of data is going to be
orders of magnitude larger.  Also, we release this software publicly under GNU GPL
License\footnote{\url{http://www.gnu.org/licenses/gpl.html}}, so that it may be studied and used by
other scientists.  Therefore we have reviewed many of the available technologies and chosen those
most promising to be applied to large datasets.  We focused in particular to the most standard and
robust technologies, favoring simplicity and maintainability over pure performance.

\subsection{Data storage}

{\tt HDF5} is the most widespread file format for new applications in the scientific community
\citep{hdf5}, as it allows binary machine independent hierarchical data to be stored efficiently on
disk and supports parallel reading and writing through {\tt MPI}\citep{mpi04}. Nowadays the {\tt
HDF5} {\tt C}, {\tt C++}, {\tt Fortran} libraries are installed in nearly every supercomputer, and
can also easily be installed on desktops.  {\tt HDF5} software is definitely complex, because it
allows extremely flexible data selection and format conversion of multi-dimensional datasets, but
still it is one of the easiest options for high performance parallel I/O.

\subsection{Parallel linear algebra}

The key element for the implementation of the destriper is the choice of a distributed linear
algebra package, the most popular packages in the scientific community are {\tt
PETSc}\footnote{\url{http://www.mcs.anl.gov/petsc/}} by Argonne National Laboratories \citep{petsc}
and {\tt Trilinos}\footnote{\url{http://trilinos.sandia.gov/}} by Sandia National Laboratories
\citep{trilinos}.  Both are implemented in {\tt C++} and provide implementation of distributed
arrays through MPI and several solvers for linear and non-linear systems.  We favored {\tt
Trilinos} because the APIs are more modular, fully Object Oriented, therefore easier to work with.

\subsection{Programming language}

{\tt Trilinos} is implemented in {\tt C++} and has a native {\tt Python} interface named {\tt
PyTrilinos}, built on top of {\tt NumPy}. {\tt PyTrilinos} offers an interface to the underlying
{\tt C++} objects with minor performance degradation, as both the communication and the math
operations are executed by the optimized {\tt C++} library.  We implemented the {\tt Python}
destriper {\tt dst}, which implements data distribution using {\tt Epetra} (Trilinos package), data
loading using {\tt HDF5}, GMRES solver using {\tt Belos} ({\tt Trilinos} package), and {\tt
cython}\footnote{\url{http://www.cython.org}} for two computationally intensive functions.

\subsection{MPI communication strategies}

Data, baselines and maps are distributed uniformly among the available processors, each process has
a local map which involves just the pixels (HEALPix) hit by its own section of the data, therefore
the same pixels are duplicated in several processes.  When the destriping algorithm requires the
application of the matrix $P^T$ first the data are binned locally on each process, then the local
maps are distributed and summed to the global map which has unique pixels and that is distributed
uniformly among the processes, independently from the local map.  This process is automatically
performed by {\tt Trilinos} with highly optimized communication strategies, it is just necessary to
specify the source and target distributions.  Typically the map is weighted and then rescanned to
time domain, therefore the opposite operation is performed, where the pixel values from the global
unique map are distributed to all the local maps and overwrite the previous value. Then rescanning
to time domain is simply a local operation.

\subsection{Access to the code}

The full source code is available on github, \url{http://github.com/zonca/dst}, under GNU GPL Free Software 
license. The software is designed to destripe any dataset in {\tt HDF5} format which provides
$I$, $q$ and $u$ data and pointing information in the correct format, and outputs hitmaps, binned
maps, destriped maps and baseline arrays.

\section{Simulations}
\label{sec:simulations}

All maps in the next sections are projected with a Mollweide projection. Since B-Machine covers about
50\% of the sky, we have sliced the full sky view focusing on the observed area, a full sky image
in Galactic coordinates is available in Fig.~\ref{fig:ch6tdestripedgalactic}.

\begin{figure}[htb] \includegraphics[width=\columnwidth]{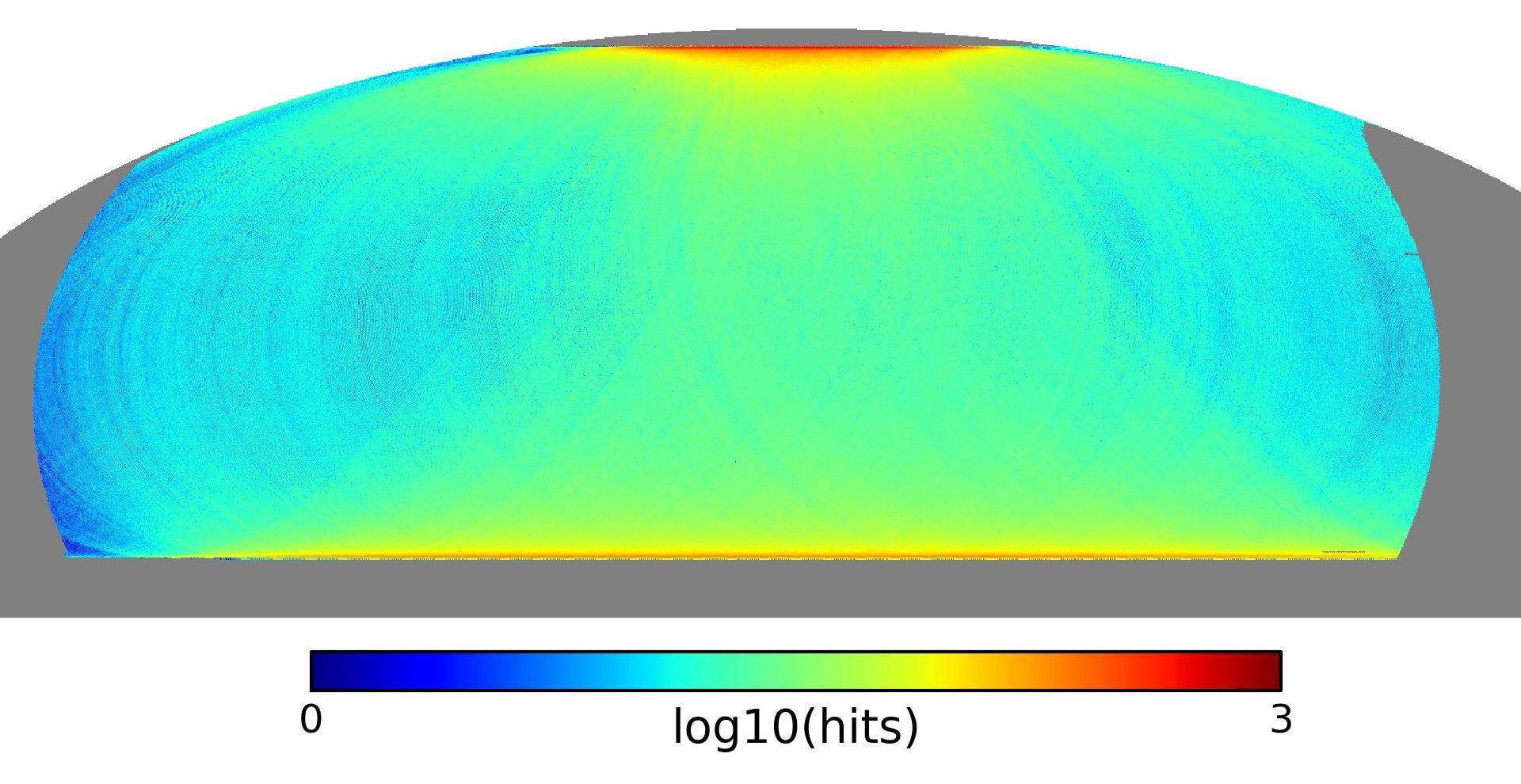} 
    \caption{Base 10 logarithm of the hitcount map for B-Machine
    channel 6 in Equatorial coordinates at HEALPix $N_{side}$ 512, i.e.\ a pixel area of 47.2 square 
arcminutes for about 30 days of observing time.\label{fig:hitmap}}
\end{figure}

\begin{figure}[htb] \includegraphics[width=\columnwidth]{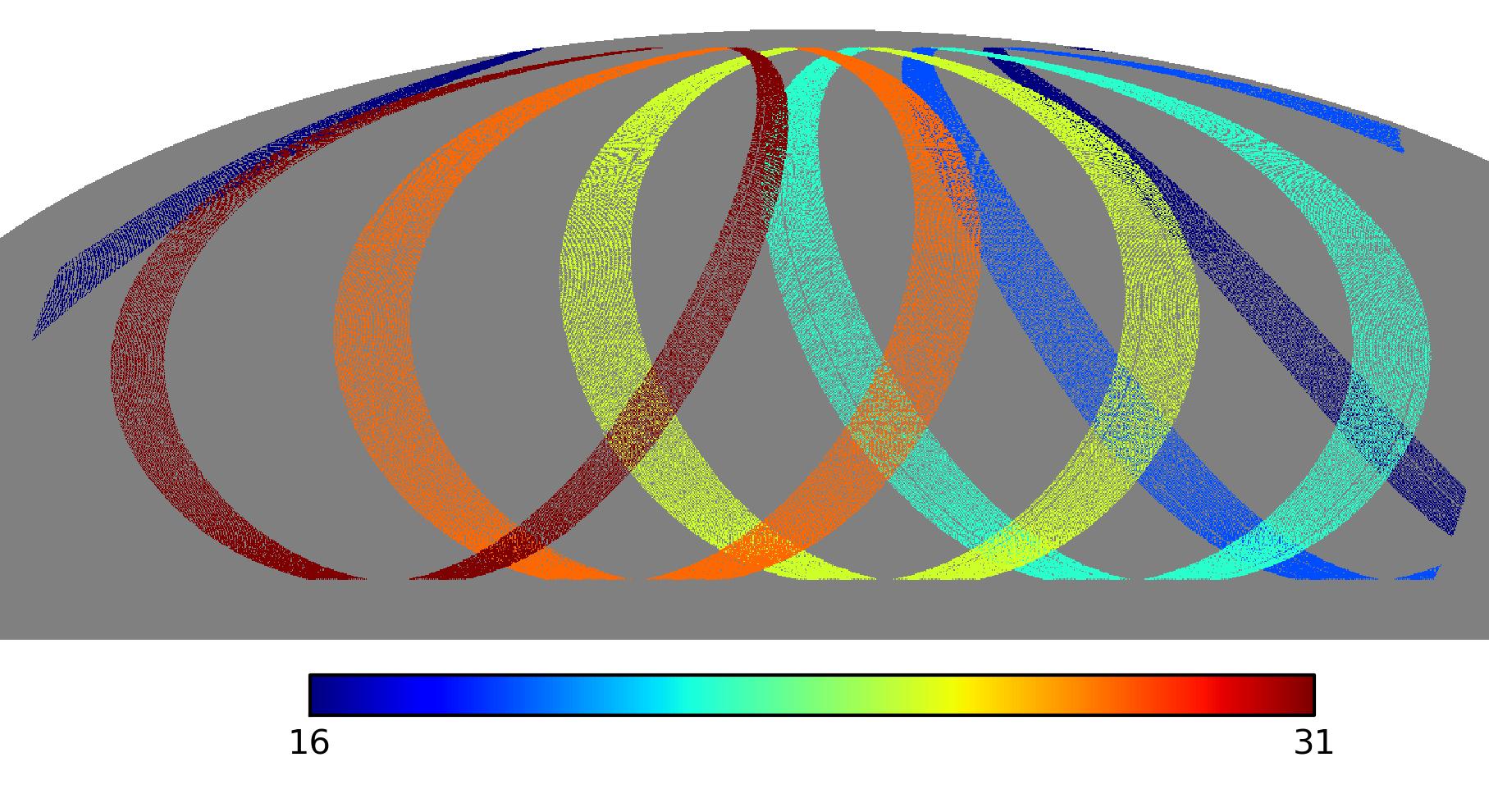} 
    \caption{B-Machine scanning strategy for one night of acquisition: 
        each ring displays the pixels covered by one hour of data, 
        about 51 revolutions at 70 seconds per revolution, from 16, i.e.\ 
        hits between 16:00 and 17:00, to 31, i.e.\ hits between 7:00 and 8:00,
        every 3 hours, i.e.\ the second ring is 19:00 to 20:00.
        Crossings between different rings provide 
leverage to the destriping algorithm for solving the baselines.
\label{fig:circles}}
\end{figure}

We have produced a set of test data using the same pointing information of one of the B-Machine channels and
simulated timelines of growing complexity. 
Sky coverage is about 44.8\%; on a HEALPix $N_{side}$ 512 pixelization, i.e.\ resolution
of about 6.9\arcmin, the median hitcount per pixel is 20, 
with higher hitcount toward the North Pole and the Equator, see a channel hitcount map in Fig.~\ref{fig:hitmap}.

Fixed elevation scanning (45\degreenospace) at 1 revolution each 70 seconds gives good coverage and frequent
crossings near the North pole and the Equator in the Equatorial reference frame.
Fig.~\ref{fig:circles} illustrates the scanning strategy for August 9\textsuperscript{th}, 2008; 
each ring plots the covered pixels for one hour
acquisition time, we plotted a ring each 3 hours, at 16:00, 19:00, 22:00, 1:00, 4:00 and 7:00 in local time,
from right to left, i.e.\ from 16:00 in blue to 31:00 (7:00) in red.
The sections of the data nearby sunrise and sunset were contaminated by large thermally induced
gain variations and were masked out.

The input signal was based on WMAP 7 years $Q$ band, smoothed by 1\degree in order to reduce the noise in map domain; white noise variance and $1/f$ noise knee frequency and slope were based B-Machine
channel 6 noise properties.  

 The choice of the baselines length is typically related to the spinning frequency of the experiment, a
baseline length of about the length of a scanning ring helps a more uniform distribution of the
crossing points, which are the key to a successful solution of the destriping equations.  B-Machine
spins around its vertical axis in about 70 seconds, therefore we chose a baseline length slightly
shorter, of about 60 seconds, i.e. 2000 samples at 33.4 Hz. 

Signal only simulations show that the destriper solves for null baselines, up to the machine
precision of $10^{-16}$, as expected by the fact that the input signal has been generated itself by a
map and therefore all measurements on the same pixel have a consistent signal.  Other tests involved
white noise only or $1/f$ noise only simulations.

\begin{figure}[bth] \includegraphics[width=\columnwidth]{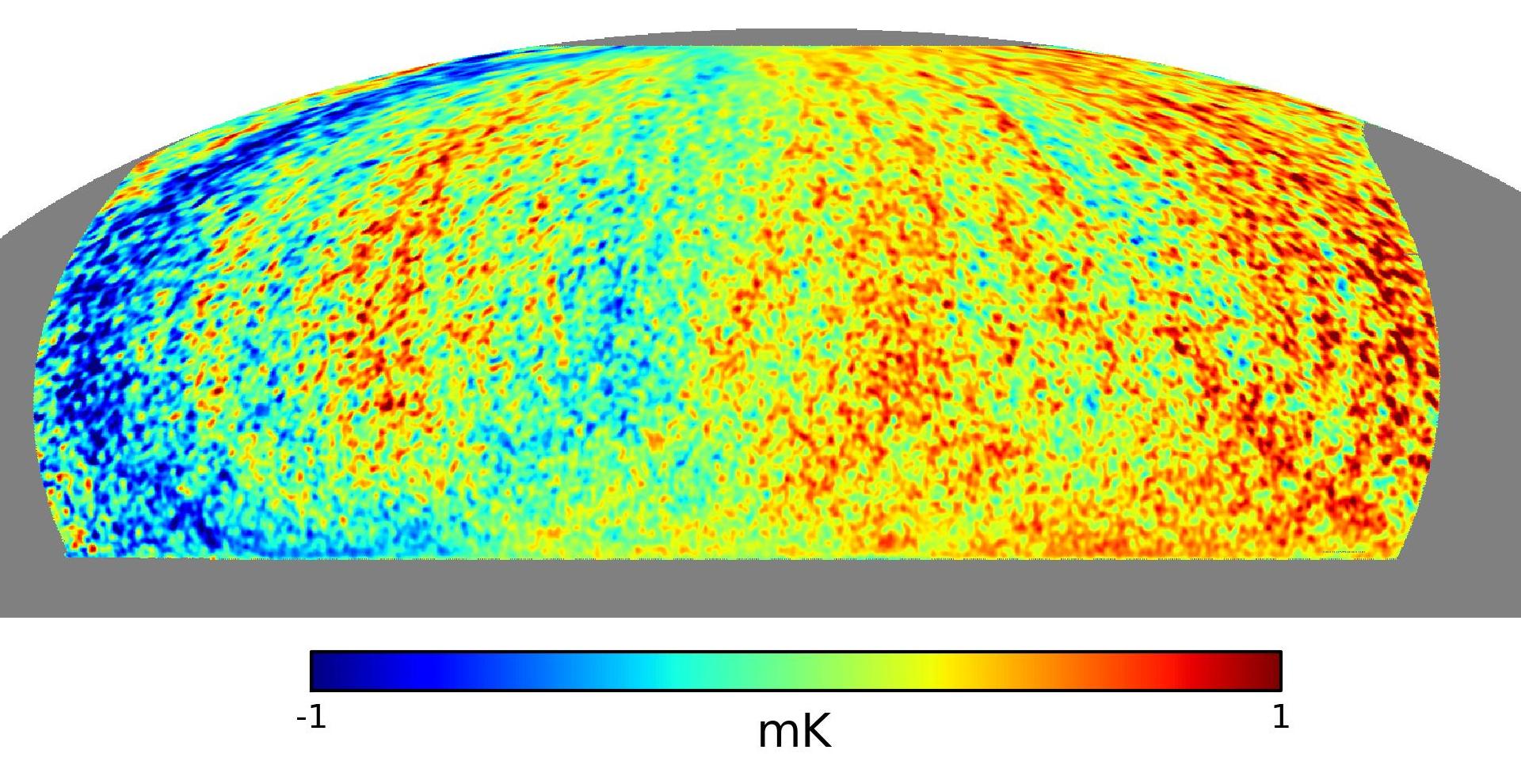}
    \includegraphics[width=\columnwidth]{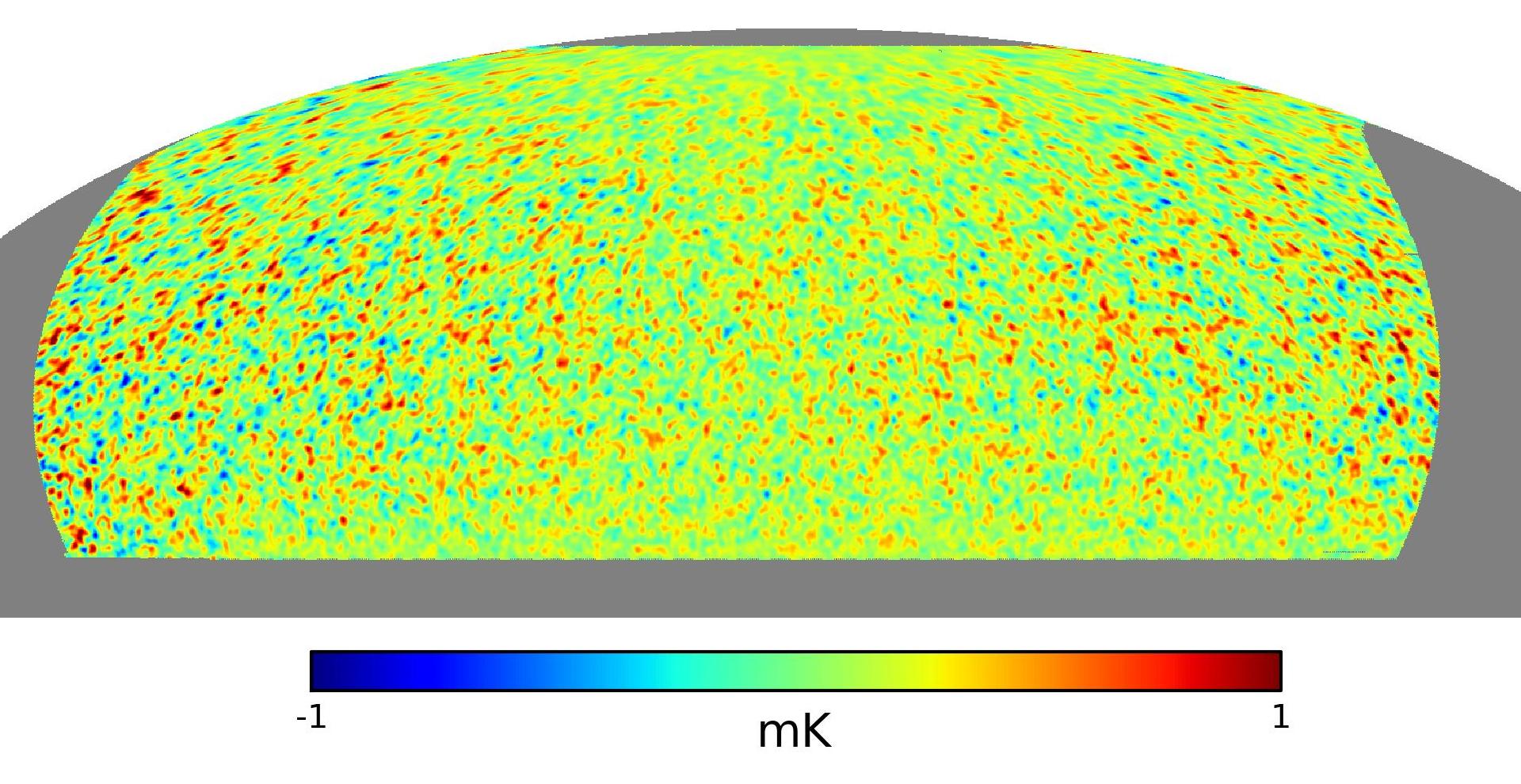} \caption{Difference between binned
        (top) and destriped (bottom) simulated maps and the input signal, smoothed with a 1\degree gaussian beam. 
        The simulations were based on WMAP signal and simulated $1/f$ noise.
        Destriping removes the stripes
caused by correlated noise, residuals are consistent with white noise} 
\label{fig:simmaps} \end{figure}

The most interesting simulation involves signal, $1/f$ and white noise, in this case we have
computed the residual maps by removing the input signal from the output maps; the difference between
the input signal and the output before destriping (top of Fig.~\ref{fig:simmaps}) smoothed by 
1\degreenospace, is dominated by stripes caused by correlated noise.  After removing the baselines from the
timestream and produced the destriped maps (bottom of Fig.~\ref{fig:simmaps}), the measurement is
not affected by any stripe, but dominated by white noise.  The amplitude of the residual white noise
is modulated by the hits per pixels and agrees with the hitmap in Fig.~\ref{fig:hitmap}.

\begin{figure}[H] \includegraphics[width=\columnwidth]{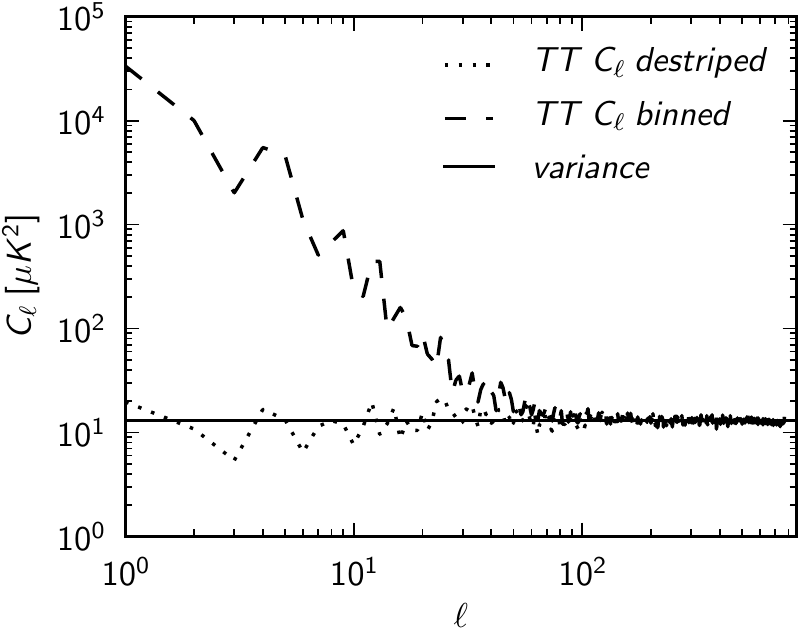}
\includegraphics[width=\columnwidth]{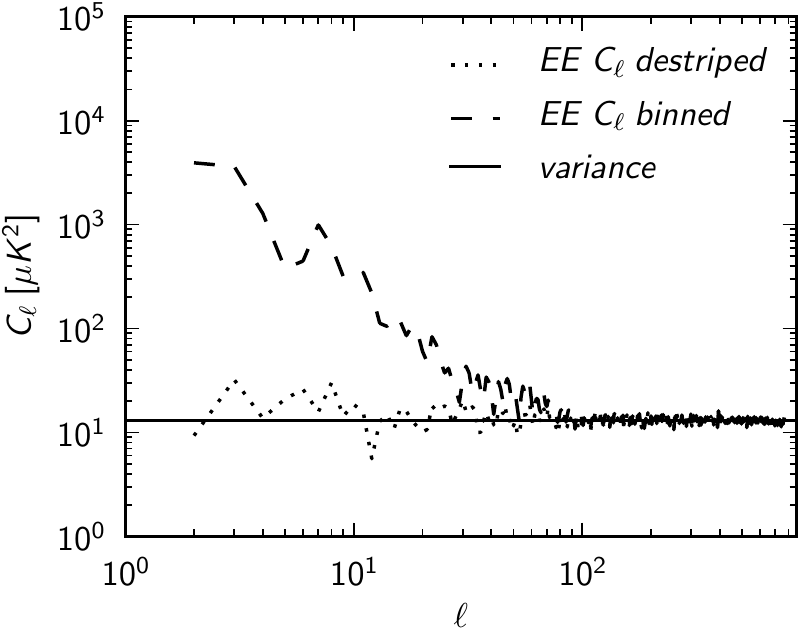} \caption{$TT$ and $EE$ angular power spectra of the noise in the
binned and destriped maps of the simulation of signal and $1/f$ noise based on B-Machine scanning strategy and
noise properties.
Destriping completely removes correlated noise both in temperature and polarization, the residual angular power spectrum after destriping agrees with the expected level computed from the variance of pure white noise.
} \label{fig:testTTEE} \end{figure}

Fig.~\ref{fig:testTTEE} shows the spherical harmonic transforms of the binned and destriped map
residuals compared to the expected noise due to the instrument white noise: destriping effectively
removes all the low frequency power due to $1/f$ signal and retrieves a measurement just dominated
by white noise, both in temperature and in polarization. The reduced $\chi^2$ of the residual map
drops from $1.052$ for $I$ and $1.026$ in $Q$ in the binned map to $1.0002$ for $I$ and $1.0012$ for $Q$
in the destriped map.

Other sources of systematic effects that cannot be modeled as $1/f$ noise with a knee frequency
lower than the spin frequency are not affected by the
destriping algorithm, and need to be dealt with before applying destriping.
In the rest of the section we will show two illustrative examples of these types of systematics effects which can affect B-Machine and
show their impact on the destriping algorithm. Since such systematics are instrument-specific and vary
largely, we do not try to be exhaustive or general.

\begin{figure}[h] \includegraphics[width=\columnwidth]{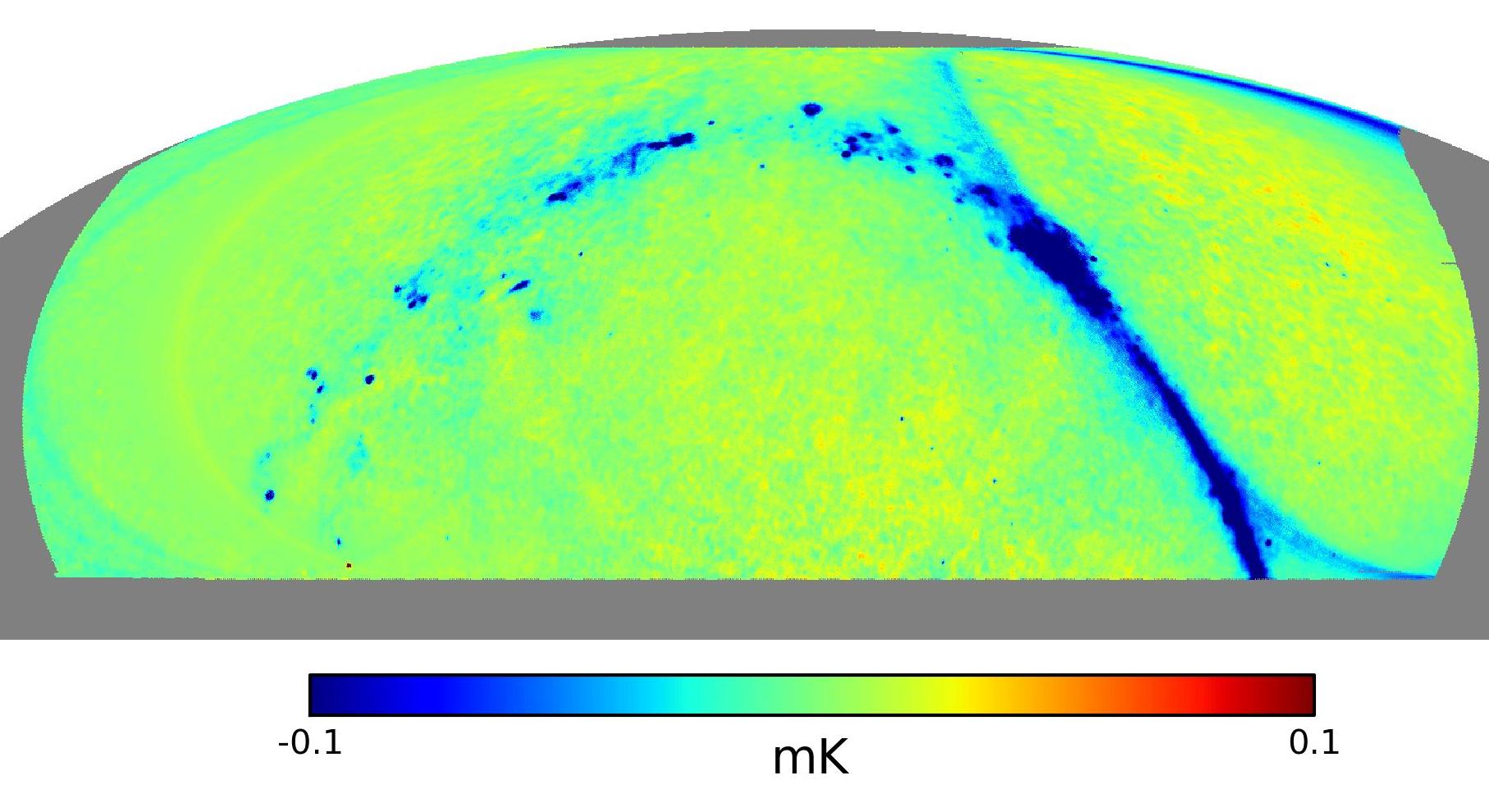} \caption{$I$ map of a signal-only simulation with
    uncorrected $10\%$ gain fluctuation on a daily timescale, the destriper solves for spurious baselines 
on the rings that cross the brightest regions of the galaxy.}
\label{fig:gaindrift} \end{figure}

Gain variations of the instrument are multiplicative effects, and are not efficiently
corrected for by a destriping algorithm. 
Actually the destriping algorithm in such scenario can be harmful, as it spreads the effect of the
uncorrected gain error to a larger area of the map.
We created a signal-only simulation with a $10\%$ uncorrected 
gain fluctuation on a day timescale, modeled as a sinusoidal effect with maximum at noon and minimum at midnight.
B-Machine observed from 5pm to 8am, therefore most of the gain error shows up as a negative residual in
the map, stronger on the galactic plane.
The destriping algorithm in this case tries to correct the gain error by solving for negative baselines,
however this mainly has the effect of creating rings of spurious signals that affect the whole sky circle
that crosses the brightest regions of the galaxy, see the top right part of Fig.~\ref{fig:gaindrift}.
Results in $Q$ and $U$ are similar, but not as dramatic due to the much lower foreground emission.
This simulation is a worst case scenario, because this effect can be mitigated by aggressively 
masking the galactic emission.

\begin{figure}[h] \includegraphics[width=\columnwidth]{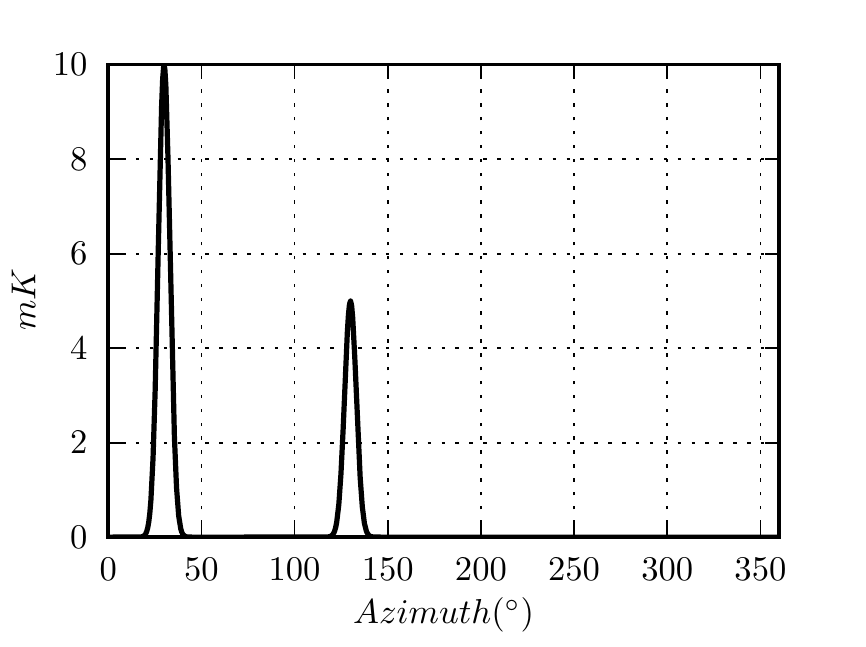} \caption{Model for a spin-synchronous RFI
    signal affecting two directions fixed in the local reference frame, 30\degree and 130\degree with a peak
height of $10$ and $5$ mK and a FWHM of $80'$}
\label{fig:spinsync} \end{figure}

\begin{figure}[h] \includegraphics[width=\columnwidth]{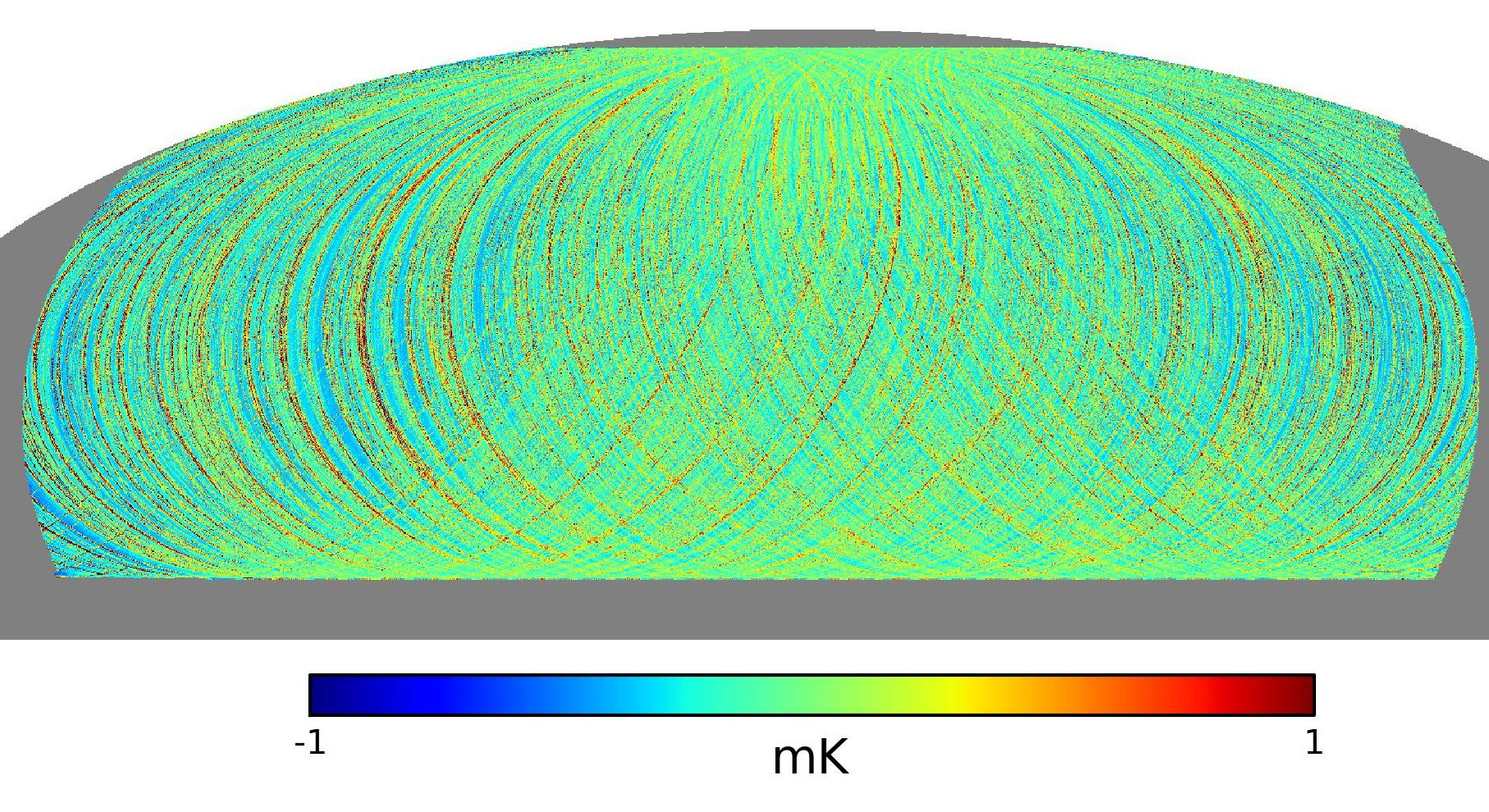} \caption{$I$ map of a signal-only simulation with
    a spin-synchronous RFI signal (model in Fig.~\ref{fig:spinsync}); the destriper is unable to remove correlated noise on timescales shorter than the baseline length}
\label{fig:spinsyncmap} \end{figure}

Another sort of systematic effects are spin-synchronous effect, typical examples for ground-based
instruments are Radio Frequency Interference (RFI) signals from ground emitters, e.g.\ radio transmitters,
radars, or satellites, e.g.\ geostationary telecommunication satellites.
RFI signals have typically a fixed azimuth, or more often 2 fixed azimuths, in the local reference frame,
when the source signal enters the telescope from directions other than the main field of view, i.e.\ 
stray-light on the often annular beam side lobes.

We created a simple model of a RFI signal with two gaussian peaks with a Full Width Half Max (FWHM) of twice the B-Machine FWHM, separated by 100\degree and with an amplitude of 10 and 5 mK, see Fig.~\ref{fig:spinsync}.
The effect of RFI in frequency domain is to add correlated noise on timescales equal and shorter than the spin frequency,
which cannot be removed by the destriper.
Both the binned and the destriped map (Fig.~\ref{fig:spinsyncmap}) show residual stripes distributed uniformly across the sky with an amplitude of the order of $1$ mK; destriping has no significant impact on the maps.
Both effects should be corrected for in time-domain before running map-making, gain variations need to be
modeled with a proper calibration process, RFI effects should be removed either using a template fit
or with convolution of the beam side lobes with the source.

\FloatBarrier
\newpage

\section{Destriping B-Machine data} 
\label{sec:bmachine}

The dataset size after cleaning is between 35 and 40 million samples for $I$, $q$ and $u$ for each
polarimeter, with the exclusion of channel 3 where hardware issues caused about 90\% of the data to
be invalid.  In the following section we present the results of map-making on channel 6 of B-Machine,
which is the best channel both for noise characteristics and impact of systematic effects. 

\subsection{Temperature maps}

\begin{figure}[thp]

\begin{subfigure}[b]{\columnwidth} \includegraphics[width=\columnwidth]{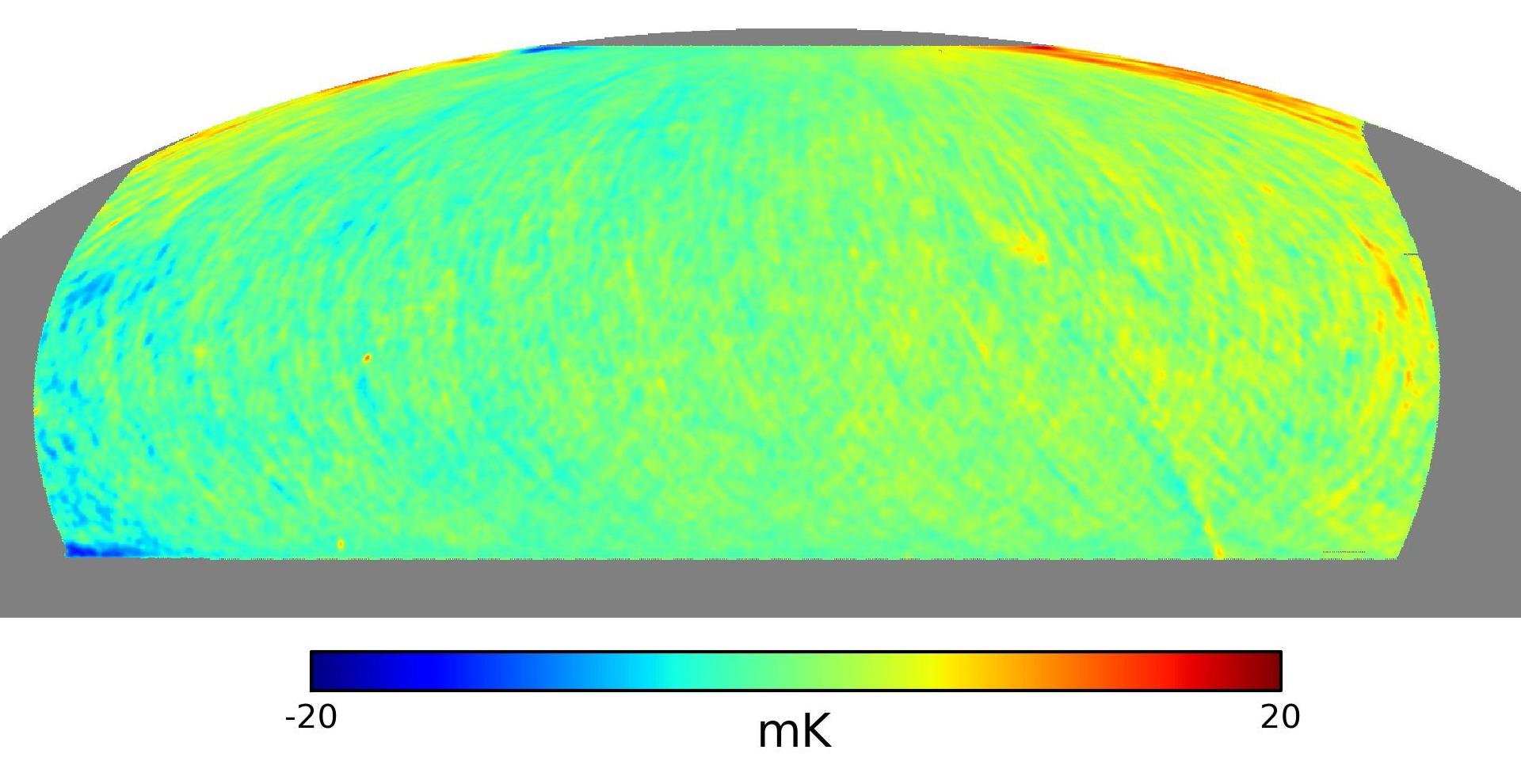}
    \caption{Binned} \label{fig:ch6tbinned} 
\end{subfigure}
    
\begin{subfigure}[b]{\columnwidth} \includegraphics[width=\columnwidth]{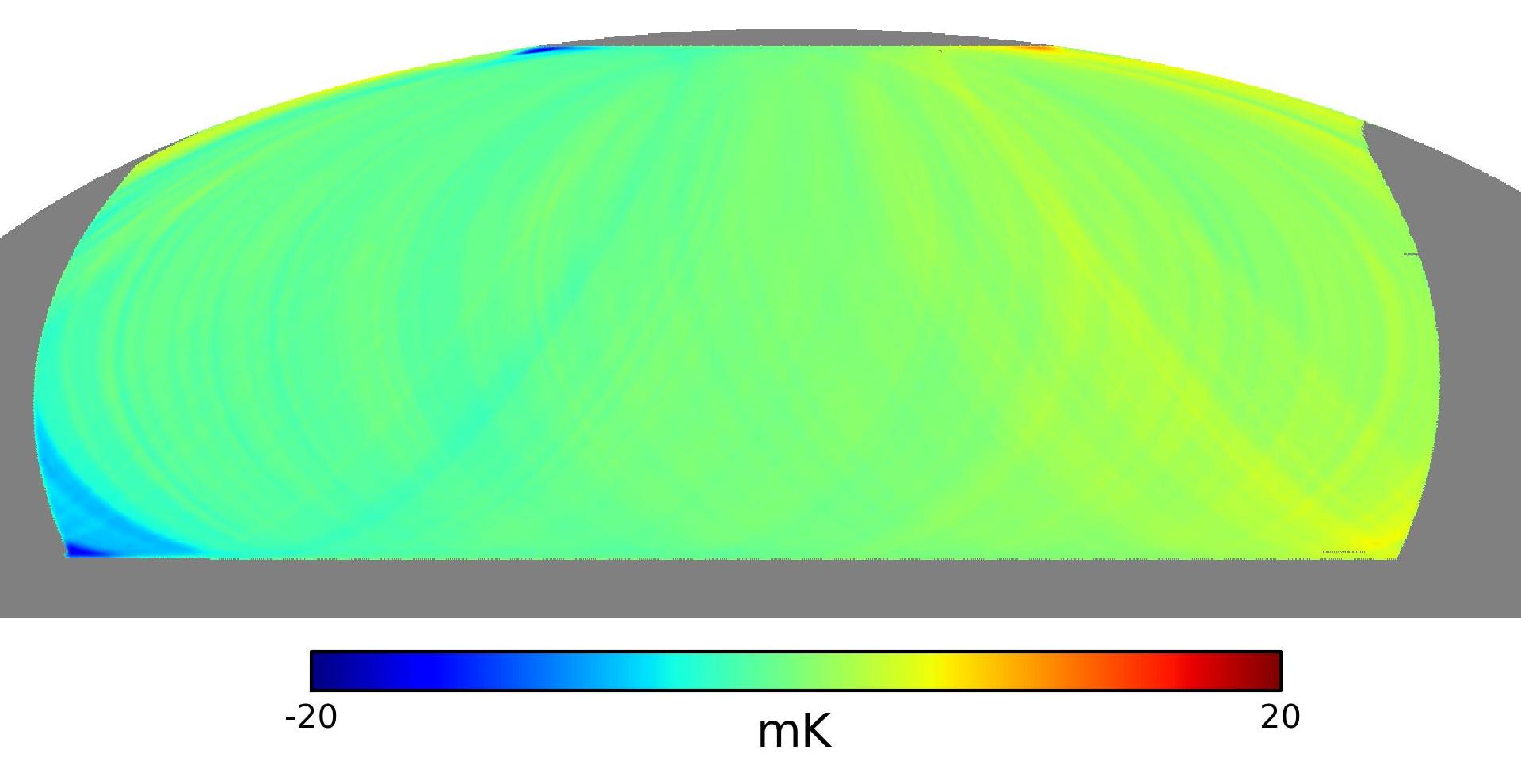}
    \caption{Baselines binned to a map, i.e.\ difference between binned and destriped}
\label{fig:ch6tbaselines} 
\end{subfigure} 

\begin{subfigure}[b]{\columnwidth}
    \includegraphics[width=\columnwidth]{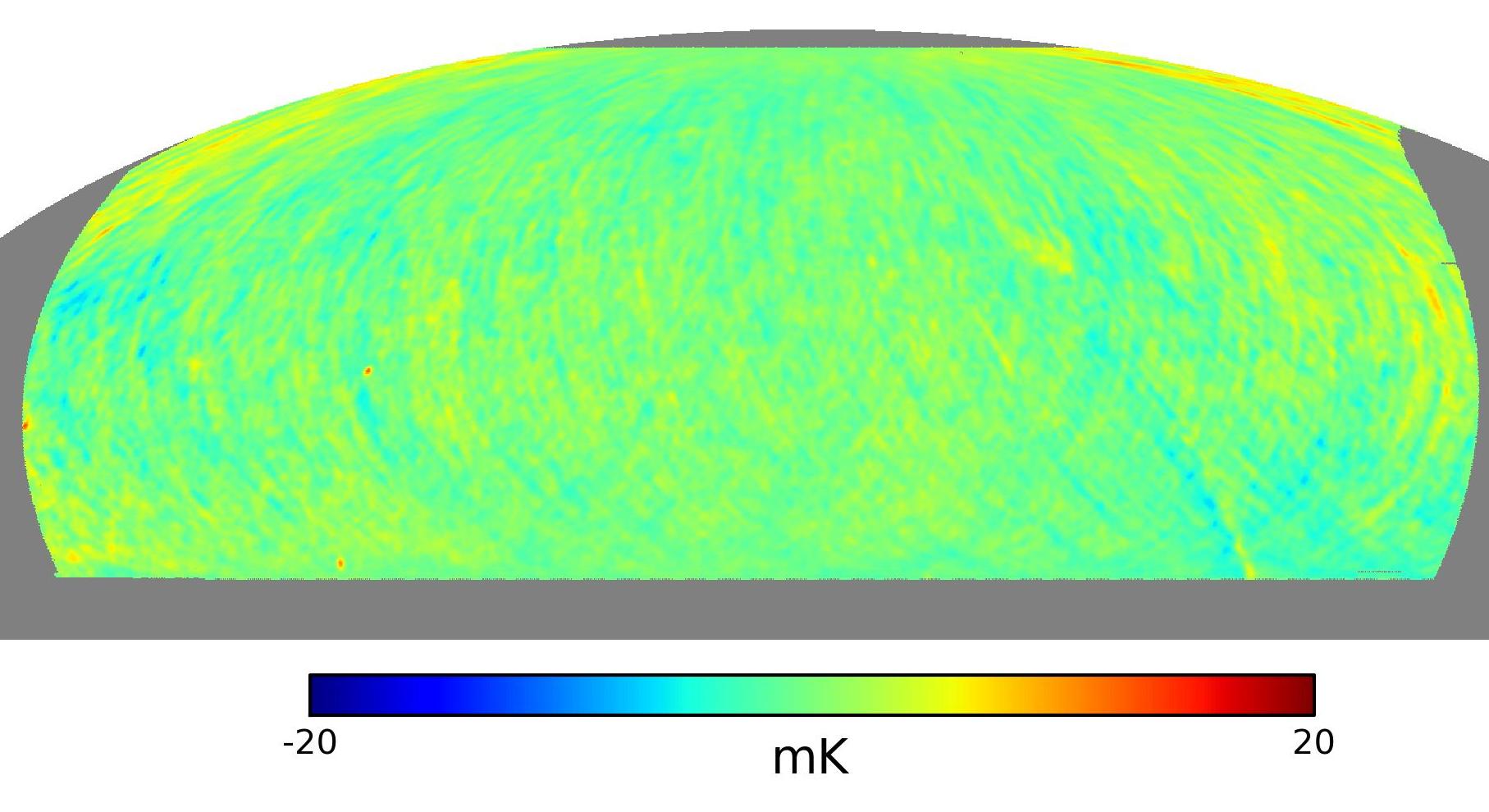} \caption{Destriped} \label{fig:ch6tdestriped}
\end{subfigure}
    
\caption{B-Machine channel 6 $I$ maps for about 30 days of acquisition in Equatorial reference frame, smoothed with a 1\degree gaussian beam. The destriping algorithm removes from the naively binned map Fig.~\ref{fig:ch6tbinned} the solved 1 minute long baselines, which can be projected to the sky as Fig.~\ref{fig:ch6tbaselines}, to achieve the destriped map Fig.~\ref{fig:ch6tdestriped}.
    Fig~\ref{fig:ch6tdestriped} in Galactic coordinates is available in Fig.~\ref{fig:ch6tdestripedgalactic}.
 }
    
\end{figure}

The B-Machine $I$ channel was not designed for scientific exploitation, it is just a secondary output
of the polarimeter, it does not benefit from the modulation/demodulation and is therefore affected by 
$1/f$ noise over the spin frequency, which cannot be efficiently removed by the destriper.

The binned $I$ map (Fig.~\ref{fig:ch6tbinned}) is dominated by significant white noise due to the
short integration time per pixel, and by large systematic effects due to temperature gradients on
the experiment.
Fig.~\ref{fig:ch6tdestriped} shows the map after destriping, and Fig.~\ref{fig:ch6tbaselines} the
difference between the binned and the destriped map, i.e.\ the baselines binned to the sky.  It is
evident that the destriping algorithm manages to remove most of the large gradients due to day-night
temperature drift.  

However, two different types of systematics are still visible in the map:

\begin{itemize}
    \item stripes all over the sky, oriented along the scanning rings are 
        likely due to the tail of $1/f$ over the spin frequency,
        which has a similar effect to the spin-synchronous RFI simulated in Sec.~\ref{sec:simulations}, Fig.\ref{fig:spinsyncmap}
    \item broader features, like the hot bands on the top of the map, are likely due to uncorrected gain drifts,
        similar to the result displayed in Fig.\ref{fig:gaindrift}.
\end{itemize}

\begin{figure*}[tb] 
    \includegraphics[width=\textwidth]{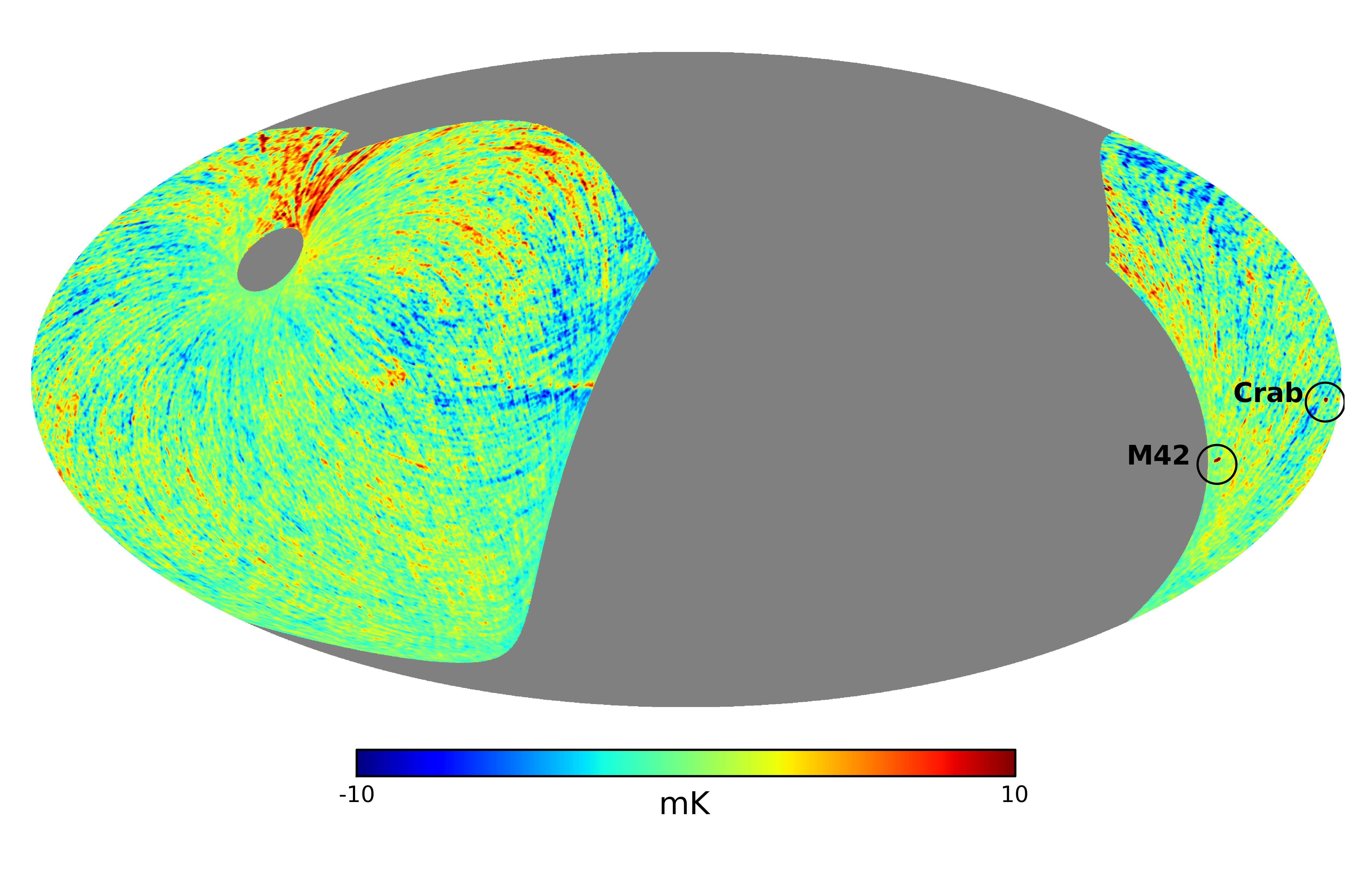} 
    \caption{Mollweide projection of the B-Machine channel 6 $I$ map after destriping in Galactic reference frame,
smoothed with a 40\arcmin gaussian beam.
Galactic emission is visible toward the center of the map, compact sources M42 and the Crab Nebula are
marked and labeled in the map.
The equivalent map in Equatorial reference frame is presented in Fig~\ref{fig:ch6tdestriped}.
    }
    \label{fig:ch6tdestripedgalactic} 
\end{figure*}

Nevertheless, Milky Way emission and at least two compact sources, M42 and the Crab Nebula are visible in the map,
see the full sky map in Galactic coordinates in Fig.~\ref{fig:ch6tdestripedgalactic} at the end of the paper. 

\subsection{Polarization maps}

\begin{figure*}[bt] 
    \begin{subfigure}[b]{\columnwidth}
\includegraphics[width=\columnwidth]{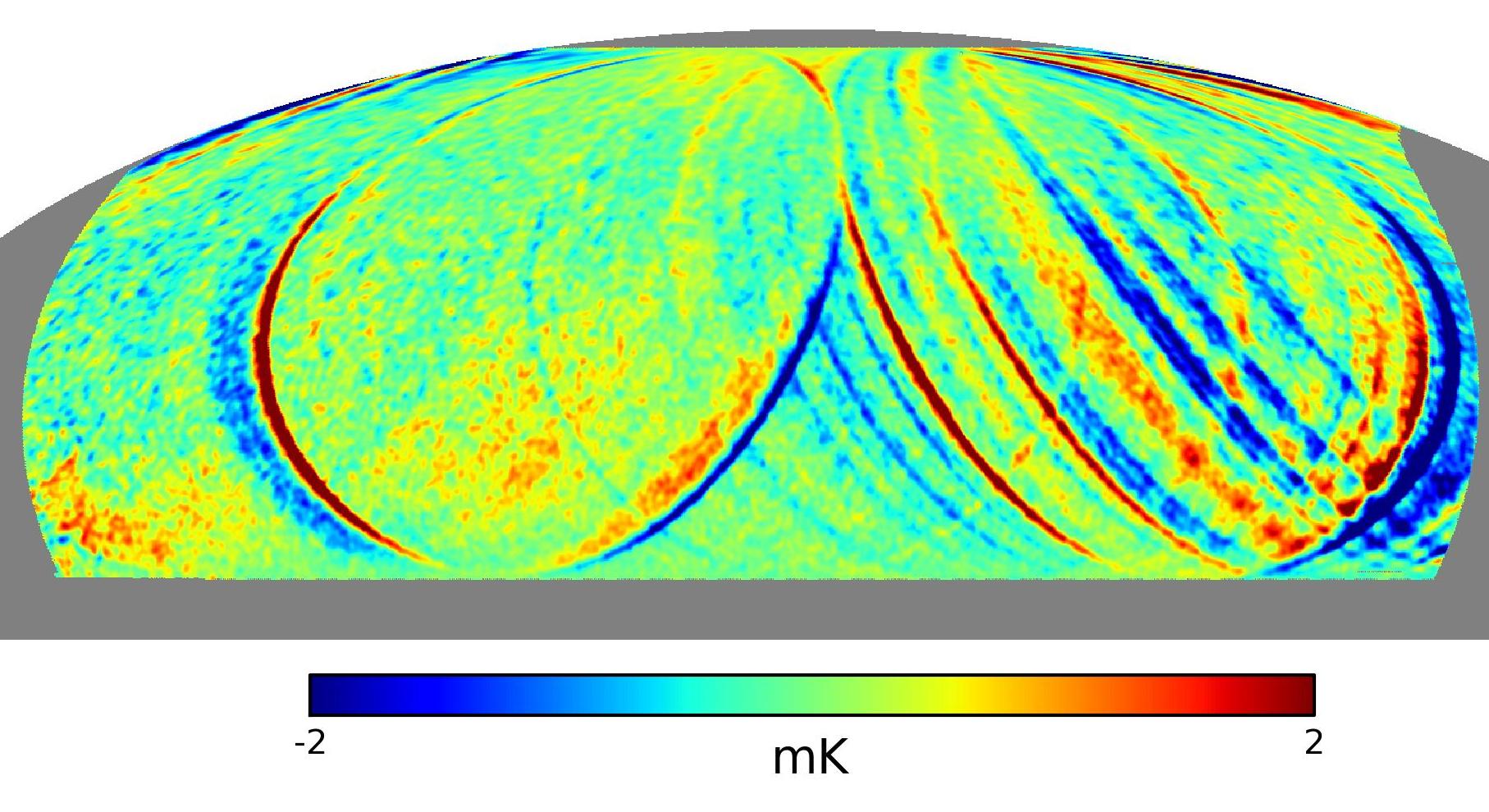} \caption{$Q$ Filtered and binned}
\label{fig:ch6q-filt}        \end{subfigure} \, 
    \begin{subfigure}[b]{\columnwidth}
\includegraphics[width=\columnwidth]{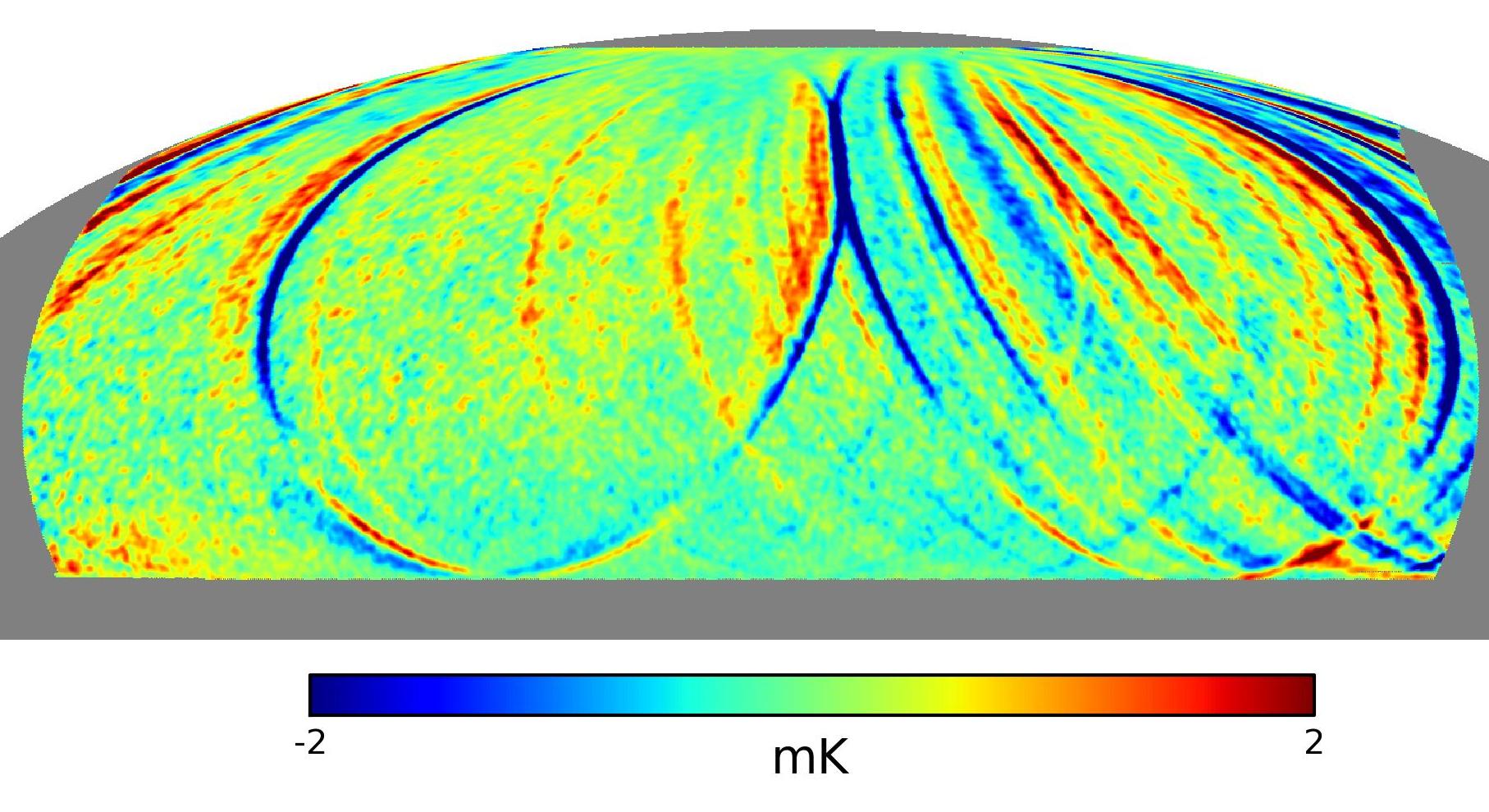} \caption{$U$ Filtered and binned}
\label{fig:ch6u-filt} \end{subfigure}

\begin{subfigure}[b]{\columnwidth} 
    \includegraphics[width=\columnwidth]{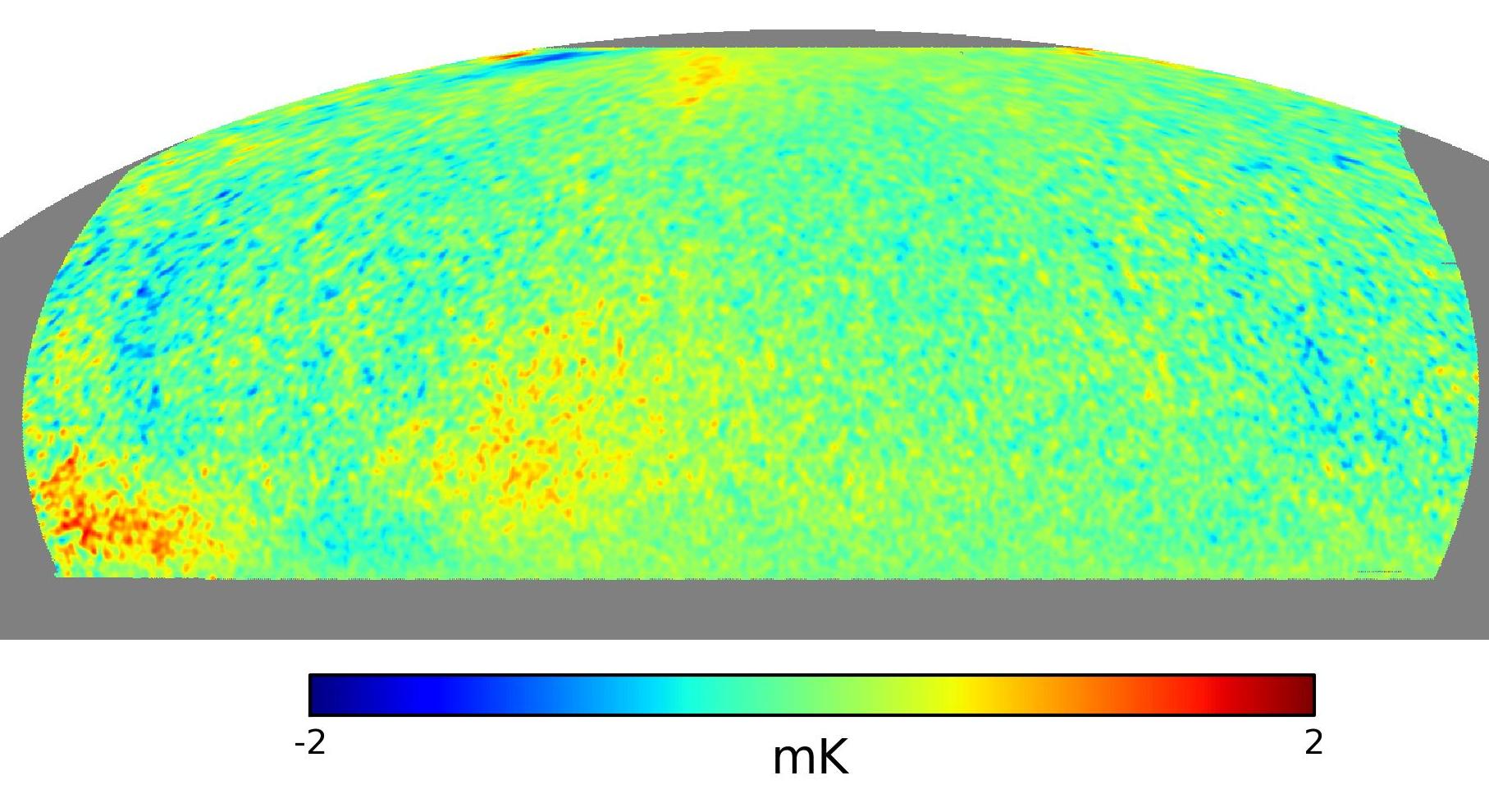}
    \caption{$Q$ Destriped} \label{fig:ch6q-dest}        \end{subfigure} \,
    \begin{subfigure}[b]{\columnwidth} \includegraphics[width=\columnwidth]{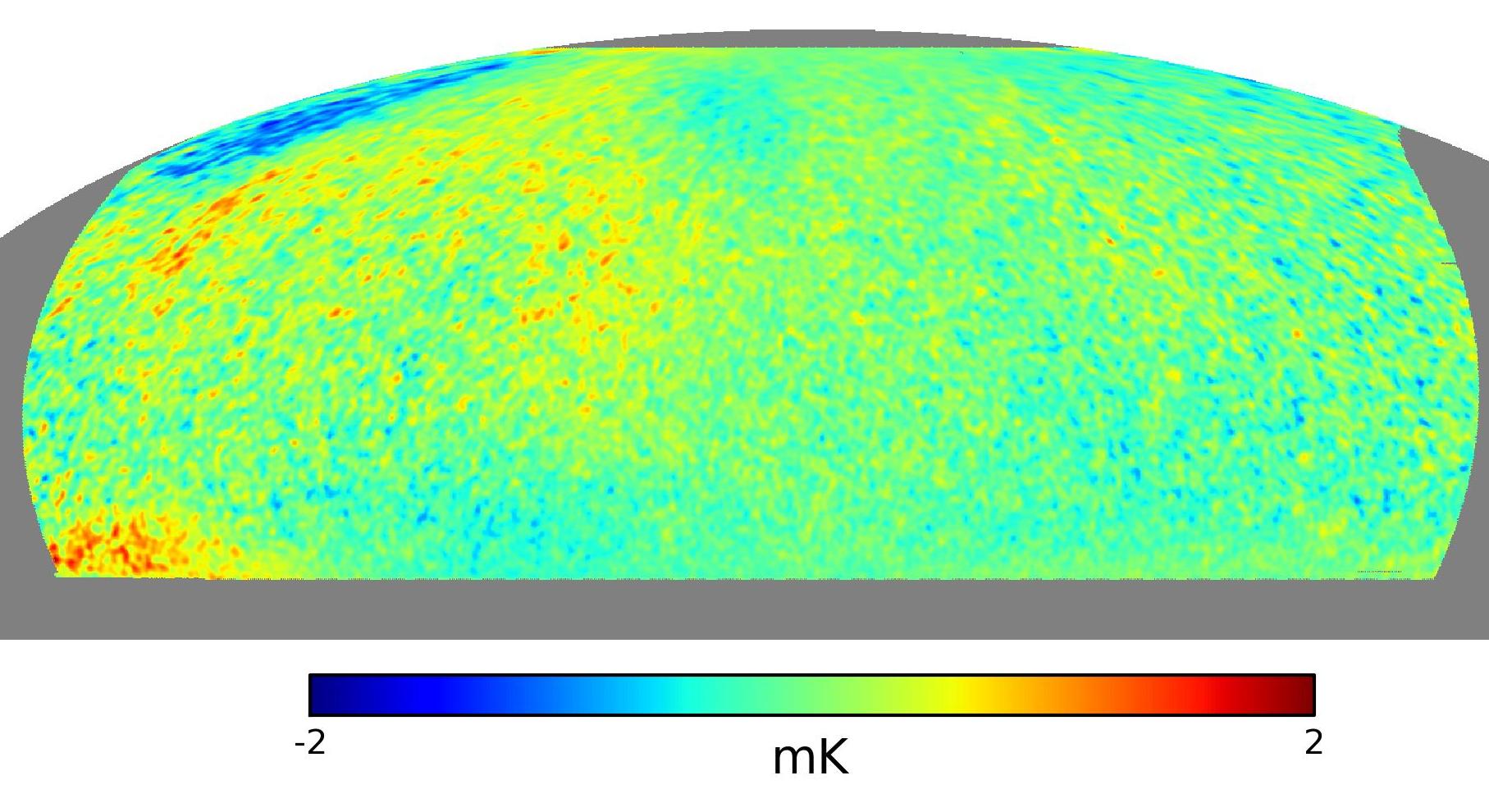}
    \caption{$U$ Destriped} \label{fig:ch6u-dest} \end{subfigure}
   
    \caption{B-Machine channel 6 $Q$ and $U$ maps in Equatorial coordinates smoothed with a 1\degree gaussian
        beam, Fig.~\ref{fig:ch6q-filt} and Fig.~\ref{fig:ch6u-filt} were produced by filtering the data with a Butterworth high
    pass filter with a cut-off at 1 hour and then directly binning to a map.
Fig.~\ref{fig:ch6q-dest} and Fig.~\ref{fig:ch6u-dest} are the results of the destriping algorithm with 1 minute long baselines.} 

\end{figure*}

The $q$ and $u$ polarimeter outputs are affected by large offsets due to the electronics that need
to be either removed by high-pass filtering the data, or they can just be dealt with by the
destriper itself, that can better constrain their value by using the crossing points.

A direct binning of polarized data would be completely dominated by their offsets, therefore a more
interesting comparison is achieved by first high-pass filtering the data with a 8\textsuperscript{th} order
Butterworth filter with a cut-off of 1 hour. The filter successfully removes the offsets but
leaves several large stripes due to thermal gradients, see Fig.~\ref{fig:ch6q-filt} and
Fig.~\ref{fig:ch6u-filt}. Those features are removed instead by the destriper, see
Fig.~\ref{fig:ch6q-dest} and Fig.~\ref{fig:ch6u-dest}.  
Moreover, filtering the data might also have an impact on the sky signal, effectively modifying the transfer function of the instrument.

However, the high white noise and few
artifacts probably due to calibration cause large scale residuals. In particular the hot blob at the
bottom left is due to gain drift caused by the rising Sun.  The white noise level in the map is
however too high to detect diffuse synchrotron emission from the Milky Way, the only visible object
is Tau A, the Crab Nebula.

Maps in HEALPix FITS format are publicly available on the Internet via
Figshare\footnote{\url{http://www.figshare.com}}
\citep{bmachinefigshare}\footnote{\url{http://dx.doi.org/10.6084/m9.figshare.644507}}. A sample
timeline of 3 days is available on the same data repository, access to the full timelines is
available upon request.

\subsection{Point sources}

\begin{figure*}[bt] 
    \begin{subfigure}{4.3cm} 
        \includegraphics[width=4.3cm]{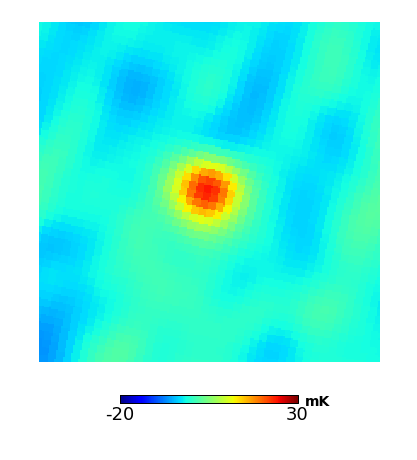}
        \caption{Tau A $I$}\label{fig:tauaI}
    \end{subfigure} \, 
    \begin{subfigure}{4.3cm}
        \includegraphics[width=4.3cm]{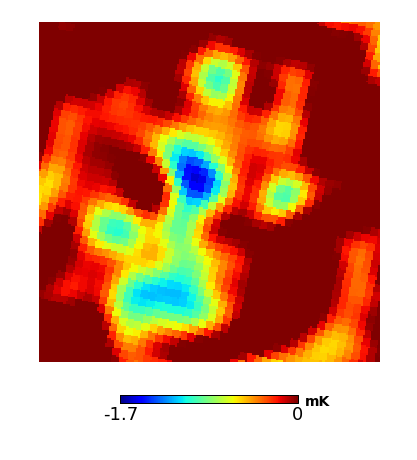} 
        \caption{Tau A $Q$}\label{fig:tauaQ}
    \end{subfigure} \,
    \begin{subfigure}{4.3cm} 
        \includegraphics[width=4.3cm]{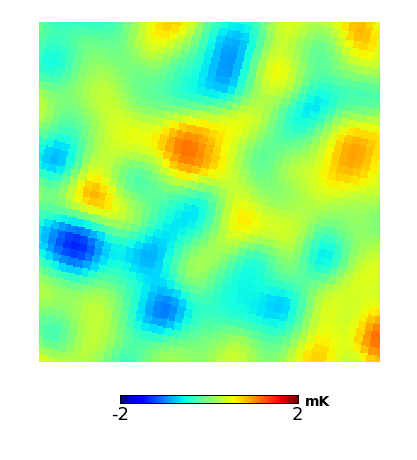} 
        \caption{Tau A $U$}\label{fig:tauaU}
    \end{subfigure} 

    \begin{subfigure}{4.3cm} 
        \includegraphics[width=4.3cm]{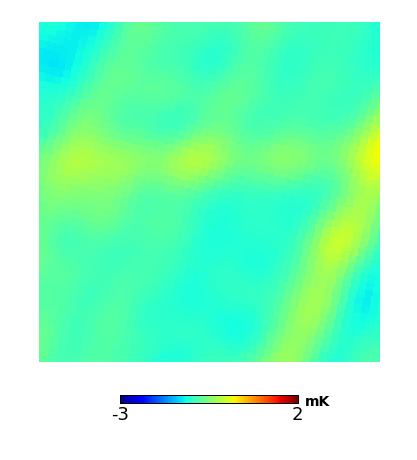}
        \caption{Tau A baselines $I$} \label{fig:tauadiffI}
    \end{subfigure} \, 
    \begin{subfigure}{4.3cm}
        \includegraphics[width=4.3cm]{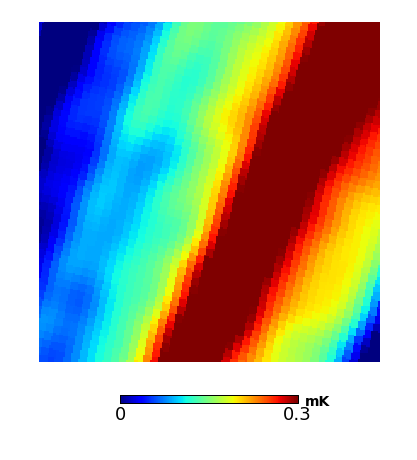} 
        \caption{Tau A baselines $Q$} \label{fig:tauadiffQ}
    \end{subfigure} \,
    \begin{subfigure}{4.3cm} 
        \includegraphics[width=4.3cm]{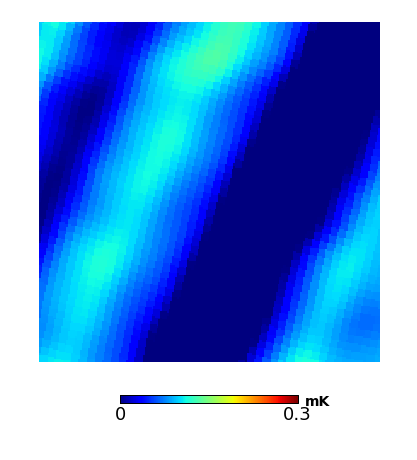} 
        \caption{Tau A baselines $U$} \label{fig:tauadiffU}
    \end{subfigure} 
    \caption{Tau A (Crab Nebula) $N_{side}$ 512 maps for channel 6 smoothed
with 40\arcmin \, gaussian beam
, \textbf{top row}: $IQU$ maps, emission in $I$ and $Q$ agrees with WMAP, emission in $U$ is lower and is not detected in the B-Machine maps.
\textbf{Bottom row}: difference between binned maps (after filtering) and
destriped maps.
Destriping in $I$ has a negligible effect on strong sources, while in polarization it is essential to
be able to identify the sources.
} \label{fig:taua} \end{figure*}

Tau A (Crab Nebula) is the only point source that is visible in the $Q$ destriped map, and it has
been used for calibration and pointing reconstruction purposes. Its flux 
agrees with WMAP Q-band \citep{weiland11} at the level of 20\% in $I$ and 6\% in $Q$. 
Tau A emission in $U$ is instead too faint to be detected in the data.

Fig.~\ref{fig:tauaI}, Fig.~\ref{fig:tauaQ} and Fig.~\ref{fig:tauaU} show $IQU$ gnomonic projection 5\degree patches around Tau A of destriped maps at
$N_{side}$ 512 and smoothed with a gaussian beam of 40\arcmin. Fig.~\ref{fig:tauadiffI}, Fig.~\ref{fig:tauadiffQ} and Fig.~\ref{fig:tauadiffU} show the
difference between binned maps (after high-pass filtering) and destriped maps.  The destriping
process has a negligible impact on $I$ maps, because the source emission is over 20 times higher
than the residual $1/f$. 
In the $Q$ polarization map, destriping is efficiently removing stripes due to correlated noise 
of the same order of magnitude of the source flux.

\FloatBarrier

\subsection{Noise characterization}

In the following section we will review the impact of destriping on the properties of the noise in the data in frequency and angular power spectrum domain.

\begin{figure}[hp] 
    \begin{subfigure}[b]{\columnwidth}
        \includegraphics[width=\columnwidth]{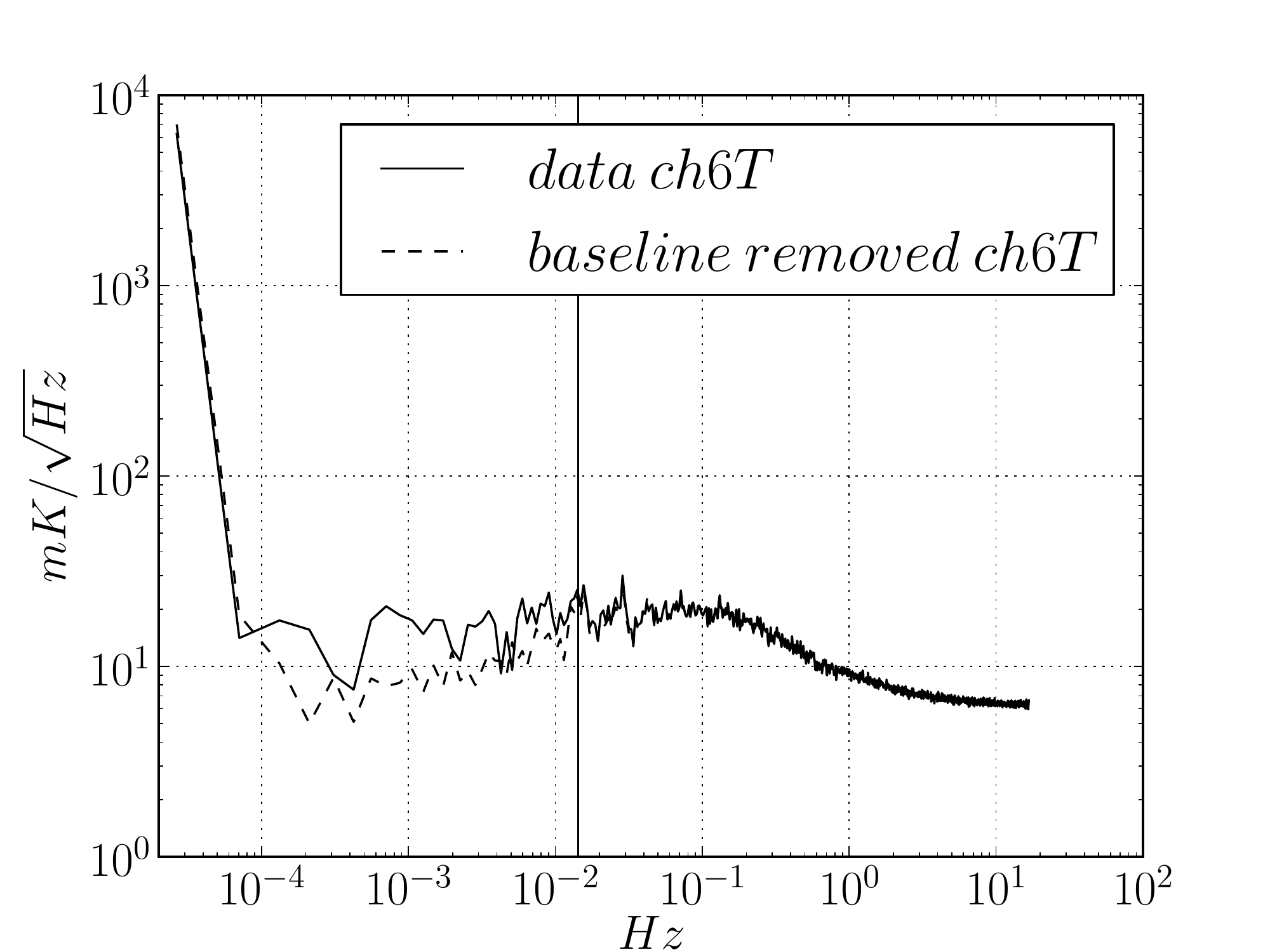} 
    \end{subfigure}
    \begin{subfigure}[b]{\columnwidth} 
        \includegraphics[width=\columnwidth]{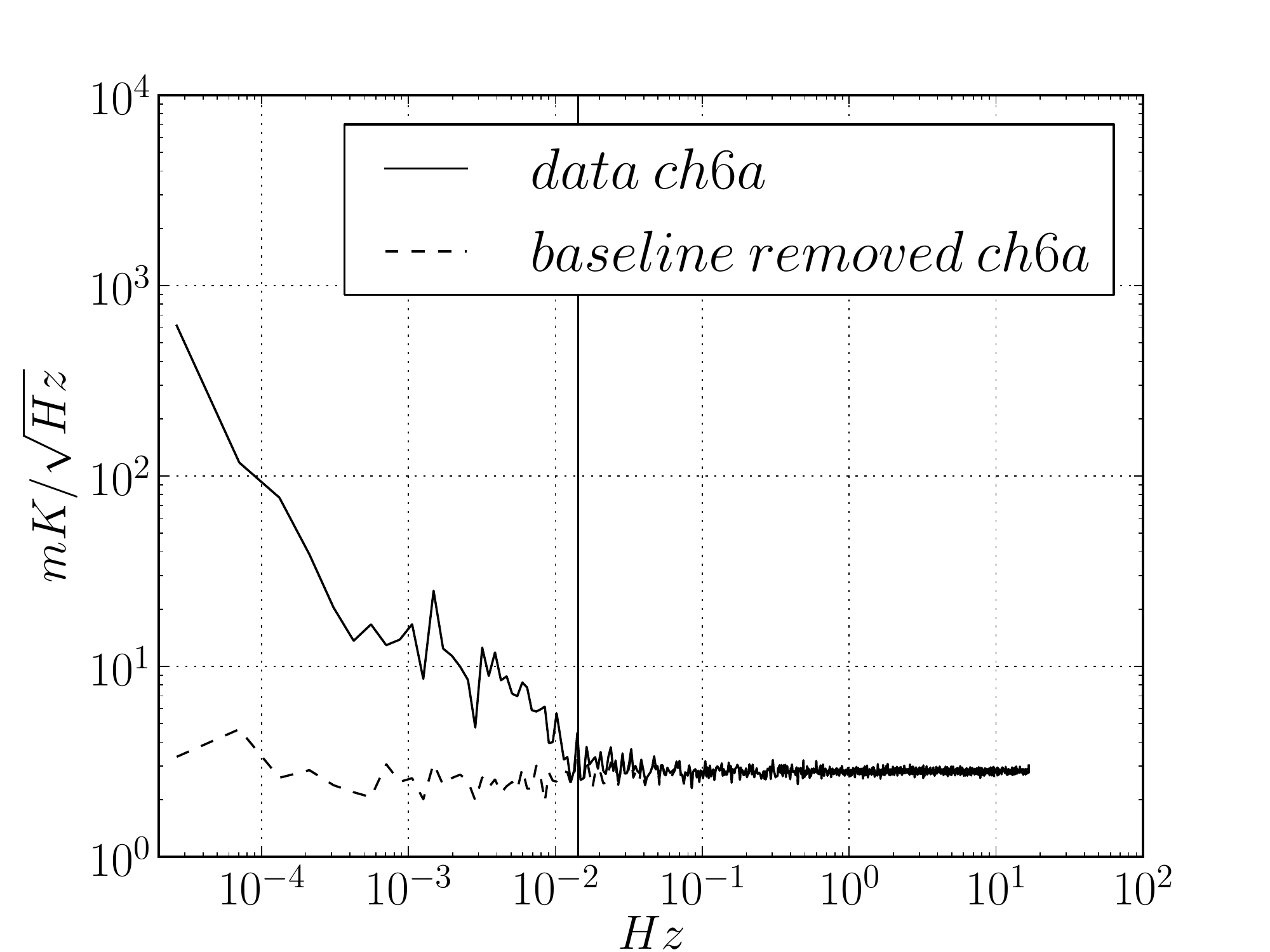}
    \end{subfigure}
    \begin{subfigure}[b]{\columnwidth}
        \includegraphics[width=\columnwidth]{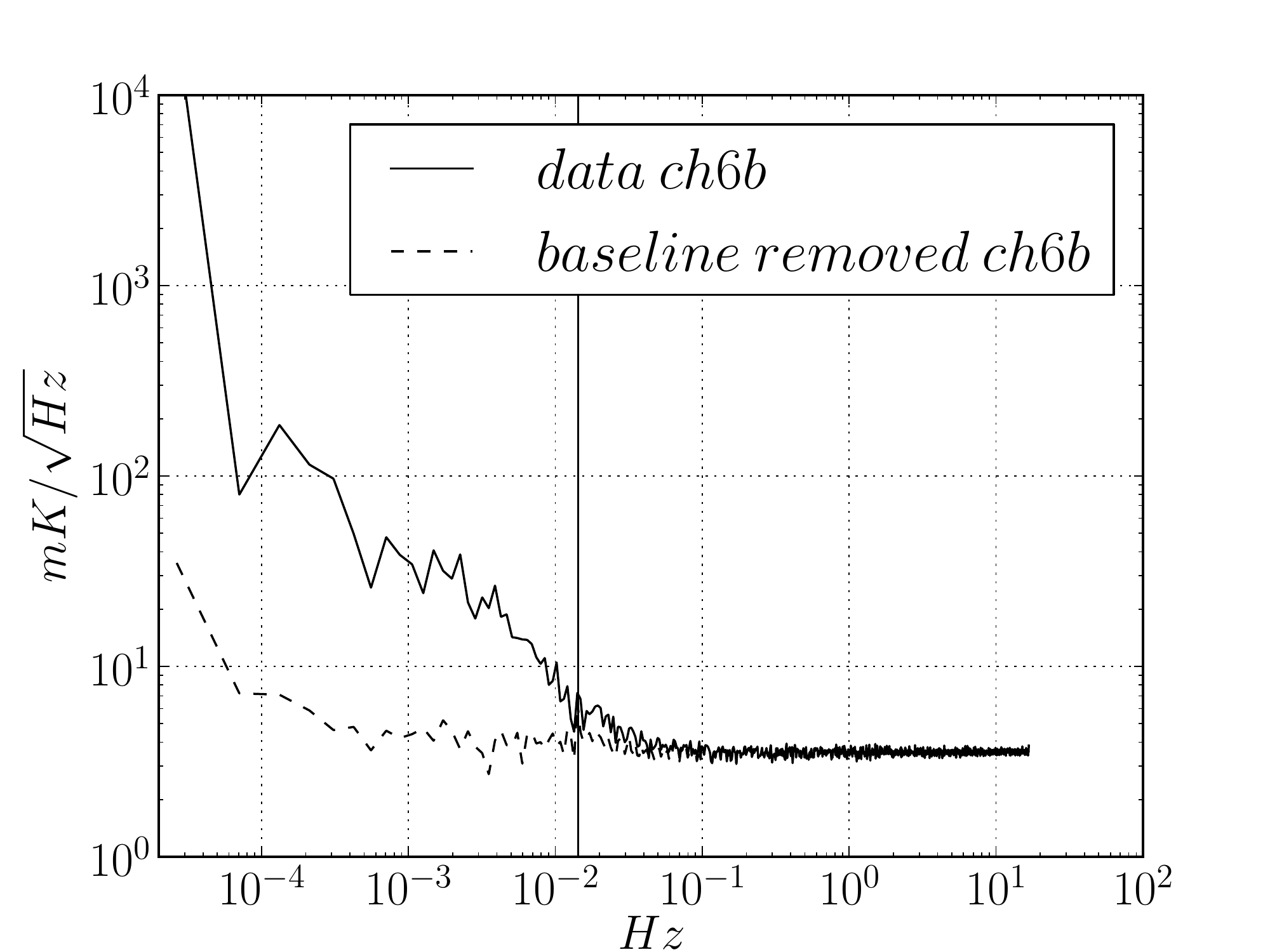} 
    \end{subfigure}  
    \caption{Channel 6 $I$, $q$ ({\tt ch6a}) and $u$ ({\tt ch6b}) data amplitude spectra before and after destriping for 1 day of data; the spectrum has been averaged over bins of linearly increasing length.} \label{fig:freq} 
\end{figure}

\subsubsection{Frequency domain}

B-Machine has very different noise properties for the temperature and the polarization channels; $q$
and $u$ channels have a low knee frequency thanks to the 120 Hz chopping provided by the
Polarization rotator, while the $I$ channel has significant correlated noise over the spinning
frequency ($\sim$1/70 Hz).

Fig.~\ref{fig:freq} shows the effect of destriping on the timelines, i.e.\ the comparison between the
signal $y$ and the baseline removed signal $y - Fa$, the $1/f$ noise below the baseline length of 60
seconds is correctly removed in the polarized channels.  In the temperature channel, instead, the
high tail of the spectrum still suffers from correlated noise, i.e.\ does not show the white noise
plateau; moreover, at low frequencies, the shape of the noise does not agree with a $1/f$ spectrum
that is one of the assumptions for the design of the destriping algorithm.
Still it is useful to apply destriping to the data to remove slower correlated noise from the timelines.

This is an extreme case for the application of a map-making algorithm, typically building an instrument
with no correlated noise on timescales shorter than the spinning frequency is a strong design driver,
which pushes either to increase the spinning frequency or more often employ strategies to improve the
noise properties of the detectors, e.g.\ fast-switching with a reference target.
In the B-Machine case the focus was only on polarization, so the $I$ channel was not designed for
scientific purposes.

\subsubsection{Angular power spectrum}

Analyzing the angular power spectrum of the destriped maps allows us to check the consistency of the
map noise with the noise level predicted by the noise properties of the receivers.  Given a white
noise standard deviation of $\sigma$, we can estimate the expected noise level in the angular power
spectrum as: 

\begin{equation} 
    C_\ell = \Omega_{\rm pix} \langle \frac{\sigma^2}{\tau} \rangle =
    \Omega_{\rm pix} \langle \frac{\sigma^2 f_{\rm samp}}{hits} \rangle, 
\end{equation} 
where
$\Omega_{pix}$ is the pixel area and $\tau$ the integration time, i.e.\ the ratio between the sampling
frequency and the hitcount map.

\begin{figure}[h] \includegraphics[width=0.95\columnwidth]{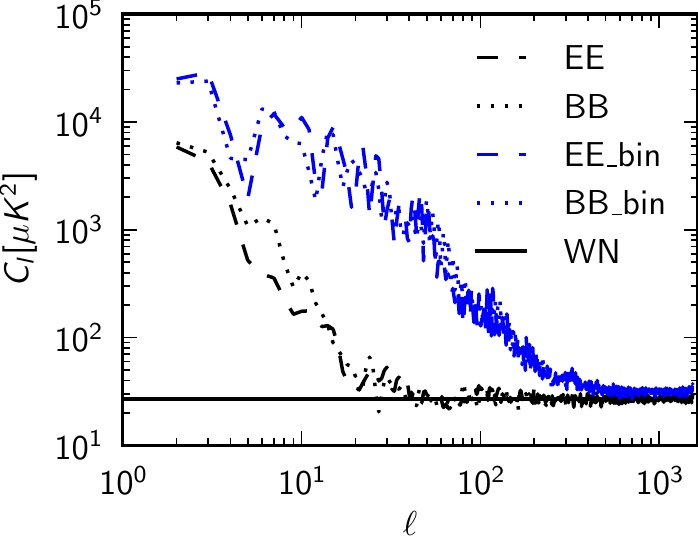} 
    \caption{Angular power spectrum
of Channel 6 $N_{side}$ 512 binned and destriped maps, corrected for the sky fraction.
We compare in blue the $EE$ and $BB$ spectra of the filtered and binned maps to the black spectra of the destriped maps, adding also as a reference a solid black line with the expected white noise floor.
Destriping removes excess power at intermediate scales  ($50<\ell<200$) due to $1/f$ noise, but cannot correct for large scale features caused by uncorrected gain fluctuations in the data.}
\label{fig:angspec} 
\end{figure}

Fig.~\ref{fig:angspec} shows the $EE$ and $BB$ angular power spectrum of the Channel 6 destriped maps extracted from the $Q$ and $U$ data, in
comparison with the binned maps of filtered data, spectra were computed on partial sky using HEALPix {\tt anafast} and then corrected by the sky fraction to achieve the equivalent spectra for a full-sky map. We just compare maps on the same sky cut, therefore we do not need a more refined analysis that corrects for the correlation across angular scales \citep{wright94}.

The large scale features in the destriped
maps after smoothing, Fig.~\ref{fig:ch6u-dest} and Fig.~\ref{fig:ch6q-dest}, strongly affect the
power spectrum at low multipoles ($\ell < 100$), even increasing the mask to about half of the available sky.
Still some residual systematic effects
impact the largest scales, probably due to gain drifts, 
but the effect of destriping is evident at intermediate scales ($50<\ell<200$) and the
power spectrum agrees with the predicted white noise variance down to $\ell = 250$.
As explained in Sec.~\ref{sec:simulations}, correcting for multiplicative effects like gain drifts 
is outside of the scope of a
destriper and the data needs to be properly calibrated before map-making.
This could not be achieved in B-Machine, but in LATTE we plan to better monitor the thermal environment
and apply a continuous calibration based on thermal sensors measurements.
Moreover, the $I$ channel in LATTE will have noise properties comparable to the polarization channels,
and we plan to use the CMB dipole to provide another calibrator for the experiment.

\newpage
\section{Test on 1 year of data}

In order to assess the scalability of {\tt dst} to larger datasets, we created a simulated dataset
with the same setup used in the validation tests for a longer time span, about 12 months of data.
We ran {\tt dst} on this dataset using the Gordon supercomputer, located in the San Diego
Supercomputing center; Gordon was designed for data-intensive applications, which is suitable for
the destriping algorithm, which has no CPU-intensive operations and is typically memory-limited.  We
configured the test run to make use of 16 MPI processes, the destriping phase required 80 GMRES
iterations to reduce the residual by a factor of $10^{12}$ and took 6 minutes.

This performance figure is a positive result considering that the software has been designed for
simplicity and robustness and not for speed. In the future, specific sections of the code can be
optimized leading to a significant boost in performance.
For example currently data are read from {\tt HDF5} files in serial mode, which is definitely going to
be the most likely bottleneck for larger applications, therefore one of the first improvements
could be in implementing a parallel version. It is going to be easy now that {\tt h5py} has native
support for the parallel version of {\tt HDF5}.

\section{Conclusion}

We have outlined a destriping algorithm tailored at polarimeters data, and described its parallel
implementation in {\tt Python} based on {\tt Trilinos} for communication and {\tt HDF5} for data
storage. We showed the results of the software on a simulated dataset and on the B-Machine 37.5 GHz
data, analyzing its impact in time-domain, in map-domain and in spherical harmonics domain.

The destriping algorithm is effective in removing correlated noise due to additive effects like
thermal fluctuations both in temperature and polarization data. In presence of other systematic effects,
like gain variations or spin-synchronous effect, destriping can be ineffective of even harmful,
they therefore need to be dealt with before the map-making stage.

The software implementation is targeted to the upcoming LATTE experiment, expected to provide about
12 months of temperature and polarization data at 8 GHz (about half billion samples), and
can easily be further optimized for larger datasets. Both the code and the data are made available
to the scientific community under Free Software licenses.

\section*{Acknowledgements} Thanks to Reijo Keskitalo for extended discussion on the formalism and
key aspects of the destriping algorithm.

This work was supported in part by National Science Foundation grants OCI-0910847, Gordon: A Data
Intensive Supercomputer; and OCI-1053575, Extreme Science and Engineering Discovery Environment
(XSEDE).

Some of the results in this paper have been derived using the HEALPix \citep{gorski05} package.

This paper has been collaboratively written using {\tt writeLaTeX}\footnote{\url{http://www.writelatex.com}}.

\bibliographystyle{elsarticle-harv} \bibliography{main}

\end{document}